\documentclass[aps,prl,superscriptaddress,nofootinbib]{revtex4}
\usepackage[dvips]{graphicx}
\usepackage{mathrsfs}
\usepackage{amsmath,amssymb}
\usepackage{hhline}
\usepackage{dcolumn}
\usepackage{bm}

\usepackage[dvips]{color}

\newcommand{\kav}{\langle k \rangle}

\newcommand{\be}{\begin{equation}}
\newcommand{\ee}{\end{equation}}
\newcommand{\bi}{\begin{itemize}}
\newcommand{\ei}{\end{itemize}}

\newcommand{\bs}{\begin{screen}}
\newcommand{\es}{\end{screen}}

\newcommand{\bssiA}{\beta_{\rm 1, S}^{{\rm SI}}}
\newcommand{\bisiA}{\beta_{\rm 1, I}^{{\rm SI}}}
\newcommand{\bssiB}{\beta_{\rm 2, S}^{{\rm SI}}}
\newcommand{\bisiB}{\beta_{\rm 2, I}^{{\rm SI}}}

\begin{document}

\title{Synergistic epidemic spreading in correlated networks}

\author{Shogo Mizutaka}
\email{mizutaka@jaist.ac.jp}
\affiliation{Japan Advanced Institute of Science and Technology, 1-1 Asahidai, Nomi 924-1292, Japan}
\author{Kizashi Mori}
\affiliation{Graduate School of Science and Engineering, Ibaraki University, 2-1-1, Bunkyo, Mito, Japan 310-8512, Japan}
\author{Takehisa Hasegawa}
\email{takehisa.hasegawa.sci@vc.ibaraki.ac.jp}
\affiliation{Graduate School of Science and Engineering, Ibaraki University, 2-1-1, Bunkyo, Mito, Japan 310-8512, Japan}

\begin{abstract} 
We investigate the effect of degree correlation on a susceptible-infected-susceptible (SIS) model with a nonlinear cooperative effect (synergy) in infectious transmissions.
In a mean-field treatment of the synergistic SIS model on a bimodal network with tunable degree correlation, we identify a discontinuous transition that is independent of the degree correlation strength unless the synergy is absent or extremely weak.
Regardless of synergy (absent or present), a positive and negative degree correlation in the model reduces and raises the epidemic threshold, respectively.
For networks with a strongly positive degree correlation, the mean-field treatment predicts the emergence of two discontinuous jumps in the steady-state infected density.
To test the mean-field treatment, we provide approximate master equations of the present model. We quantitatively confirm that the approximate master equations agree with not only all qualitative predictions of the mean-field treatment but also corresponding Monte-Carlo simulations.
\end{abstract}

\maketitle


\section{Introduction}
Infectious diseases and social contagions (the spread of information, behaviors, and attitudes) are among the most important network science topics \cite{barabasi2016network,newman2018networks}. 
The relationships between the structures of complex networks and contagion processes have been extensively reported (see reviews \cite{castellano2009statistical,pastor2015epidemic,zhang2016dynamics,de2018fundamentals,wang2019coevolution} and references therein). 
Degree inhomogeneity is known to crucially affect the spreading of an infectious disease.
Specifically, in the susceptible-infected-susceptible (SIS) model on a network where the infection transmits from an infected node to a susceptible node via their connection with an infection rate and the infected node recovers spontaneously at a recovery rate, infectious diseases can survive for a long time even with an infinitesimally small infection rate if the degree distribution $p_k$ of the network obeys $p_k \propto k^{-\gamma}$ with the degree exponent $\gamma \le 3$ \cite{pastor2001epidemic,pastor2001epidemic2}.
Subsequent works have shown that the epidemic threshold above which an SIS epidemic can persist vanishes for quenched networks with any degree exponent \cite{castellano2010thresholds,castellano2020cumulative}.

Over the past few years, an increasing body of research has considered the {\it synergistic effect} in contagion processes \cite{perez2011synergy, ludlam2012applications, taraskin2013effects, broder2015effects, gomez2016explosive, liu2017explosive, tuzon2018continuous,su2018optimal,wang2019coevolution, ogura2019synergistic, taraskin2019bifurcations}. 
The synergistic effect represents a nonlinear cooperative effect in the transmission between an infected-susceptible pair, that is, infected neighbors around them enhance the transmission rate.
This effect has been experimentally reported in biological contagions such as fungal infection in soil-borne plant pathogens \cite{ludlam2012applications} and tumor growth \cite{liotta2001microenvironment}, and in social contagions such as the spread of behaviors \cite{centola2010spread}.
Recently, St-Onge {\it et al.}\ introduced an epidemic model motivated by COVID-19, based on temporal heterogeneity of human activity, the higher-order structure of contact networks, and the minimal infective dose required for being infected \cite{st2021universal}. This model put forward the nonlinearity of  infection probability.

The role of synergy in contagions has been investigated in synergistic SIS models and susceptible-infected-removed (SIR) models on various networks; lattices, regular random graphs, random graphs, and small-world networks.
As Taraskin and P\'erez-Reche proved for the synergistic SIS model on the regular random graph \cite{taraskin2019bifurcations}, the synergistic effect can induce an {\it explosive spreading} and the emergence of a {\it hysteresis loop}.
In the classical SIS model, the infectious disease becomes extinct or persists if the infection rate is lower or higher than a specified epidemic threshold. 
The transition from the extinct regime to the endemic regime is continuous, meaning that the infected density in the steady state increases continuously from zero to some nonzero value as the infection rate increases. 
In contrast, the synergistic SIS model predicts an explosive spread of the disease with a discontinuous jump in the infected density as the infection rate increases.
The system develops a bistable region (hysteresis loop) in which the fate of the infection (extinction or endemic persistence) depends on the initial state.
Synergy in a local environment can drastically change the global spreading behaviors.
These behaviors have recently attracted interest \cite{d2019explosive}, including a COVID-19 inspired model \cite{st2021universal} and social contagions with higher-order interactions \cite{iacopini2019simplicial}.

Besides the degree inhomogeneity, the degree correlation, defined as the correlation between the degrees of directly connected nodes, is an important measure of complex networks.
Newman \cite{newman2002assortative,newman2003mixing} found that real social networks are assortative, meaning that nodes are likely to be connected to nodes of similar degrees. 
In contrast, biological and technological networks are disassortative, meaning that nodes are likely to be connected to nodes of different degrees. 
The degree correlation in complex networks also plays an important role in epidemic models 
\cite{eguiluz2002epidemic,boguna2002epidemic,boguna2003absence,moreno2003epidemic,vazquez2003resilience,van2010influence,chen2018predicting,wang2018edge,silva2019spectral,morita2021solvable}. 
For the SIS model on correlated networks, 
Bogun\'a et al. claimed that on correlated networks, the epidemic threshold of the SIS model is the inverse of the largest eigenvalue of the connectivity matrix \cite{boguna2002epidemic,boguna2003absence}. 
Following Refs.~\cite{boguna2002epidemic,boguna2003absence}, van Mieghem et al. \cite{van2010influence} reported the epidemic threshold reduces (increases) as the network becomes more assortative (disassortative). 
Very recently, Morita proposed a solvable SIS model on correlated networks and analytically showed that the epidemic threshold decreases with increasing degree correlation \cite{morita2021solvable}.
Although degree correlations discernibly affect various processes and network characteristics \cite{avalos2012assortative,jalan2016interplay,schneider2011mitigation,goltsev2008percolation,soffer2005network,serrano2005tuning,xulvi2004reshuffling}, synergistic epidemics on networks with degree correlations have been insufficiently investigated. 
Therefore, how a degree-correlated structure affects epidemic spreading combined with the synergistic effect must be discussed.

We investigate the effect of degree correlation in synergistic SIS epidemics on networks.
In this study, bimodal networks where degree correlation is tunable in a wide range are employed as degree-correlated networks.
One advantage of using bimodal networks is that it allows degree-correlated synergistic SIS epidemics to be analytically tractable.
First, we develop a mean-field treatment of the synergistic SIS to elucidate its behaviors qualitatively. 
By counting the number of fixed points of the mean-field equations, we show that unless the synergy is absent or extremely weak, the synergistic SIS model undergoes a discontinuous transition on correlated bimodal networks.
Regardless of synergy (present or absent), a positive (negative) degree correlation diminishes (enlarges) the epidemic threshold of the SIS model.
Moreover, on a strongly assortative bimodal network, the mean-field treatment predicts the emergence of two discontinuous jumps.
Next, we develop approximate master equations (AMEs) for the synergistic SIS model on correlated bimodal networks and evaluate them numerically to obtain the quantitative understanding.
The nontrivial behaviors predicted by the mean-field treatment are confirmed by the AMEs which agree with corresponding Monte-Carlo simulations.

The remainder of this paper is organized as follows. 
Section~II introduces our synergistic SIS model and a bimodal network with a tunable degree correlation. 
The first half of Sec.~III provides the mean-field treatment of the synergistic SIS model on the correlated bimodal networks, and the second half develops the AMEs of the model. 
Section~IV is devoted to a summary.

\section{Model}

This study implements a synergistic SIS model on networks.
Each node in the SIS model is in the susceptible (S) state or the infected (I) state.
In the initial state, a fraction $\rho$ of randomly selected nodes (seeds) is infected, and the other nodes are susceptible. 
The infection is transmitted to a susceptible node independently by any of its infected neighboring nodes.
A susceptible node $i$ connected to $m$ infected neighbors becomes infected with probability $\lambda_m m dt$ within an infinitesimally small time interval $dt$.
The synergistic effect is incorporated into the infection rate $\lambda_m$ of a susceptible node, which is an increasing function of the number $m$ of its infected neighbors:
\be
\lambda_m=\lambda(1+\alpha(m-1)),
\label{eq:infectionrate}
\ee
where $\lambda$ is the basic infection rate (in the absence of synergistic effect) and $\alpha$ quantifies the strength of the synergy. 
An infected node $i$ recovers to susceptible with probability $\mu dt$ within a small time interval $dt$.
Recovery is independent of the neighboring states.
The synergistic effect is constructive (enhances the disease transmission to susceptible nodes) when $\alpha>0$, and destructive (hinders the disease transmission to susceptible nodes) when $\alpha<0$.
In our analysis, we assume $\alpha \ge 0$.
When $\alpha=0$, the present model reduces to the classical SIS model.
The fraction of infected nodes in the long time limit ($t\to\infty$), i.e., the steady-state infected density $i(\infty)$, characterizes the steady-state of system. 
The system is generally either extinct with $i(\infty)=0$ or endemic with $i(\infty)>0$.
In the next section, we discuss the transition between these two states in the present model.

To understand the impact of the degree-degree correlation on the synergistic SIS epidemics, the present study employs bimodal networks which consist of two types of nodes with different degrees; 
type-$1$ nodes with degree $k_1$ and type-$2$ nodes with degree $k_2 (\le k_1)$ \cite{shiraki2010cavity,mizutaka2016robustness}.
A bimodal network takes a wide range of mixing pattern from assortative to disassortative, by adjusting the fraction of edges between nodes of different types.
It also allows us to analytically treat the model under the mean-field approximation owing to their simplicity.

We denote the fractions of type-$1$ and type-$2$ nodes by $p_1$ and $p_2(=1-p_1)$, respectively.
The degree distribution $P(k)$ of this network is given by
\be
P(k)=p_1 \delta_{k,k_1}+p_2 \delta_{k,k_2}. \label{eq:Pk}
\ee
Let $p_{xy}=P(k_y|k_x)$ be the probability that a randomly chosen neighbor of a degree-$k_x$ node has degree $k_y$ ($x,y \in \{1,2\}$).
This conditional probability $p_{xy}$ satisfies the following conditions:
\be
p_{11}+p_{12}=1,
\quad 
p_{21}+p_{22}=1,
\quad
{\rm and}
\quad
k_1 p_1 p_{12}=k_2 p_2 p_{21}.
\ee
The last condition recognizes that both $p_1 k_1 p_{12}$ and $p_2 k_2 p_{21}$ represent the number of edges between type-$1$ and type-$2$ nodes per node.
The probability $P_e(k_x,k_y)$ that the two ends of a randomly chosen edge have degrees $k_x$ and $k_y$ is given as
\be
P_e(k_x,k_y) = p_{xy} P_e(k_x),
\ee
where $P_e(k_x)$ is the probability that one end of a randomly chosen edge has degree $k_x$, i.e.,  
\be
P_e(k_x) = \frac{k_xP(k_x)}{\kav}, 
\ee
and $\kav$ is the average degree, i.e., $\kav = \sum_k k P(k) = p_1 k_1 + p_2 k_2$.

\begin{figure}[t]
\begin{center}
(a)
\includegraphics[width=.30\textwidth]{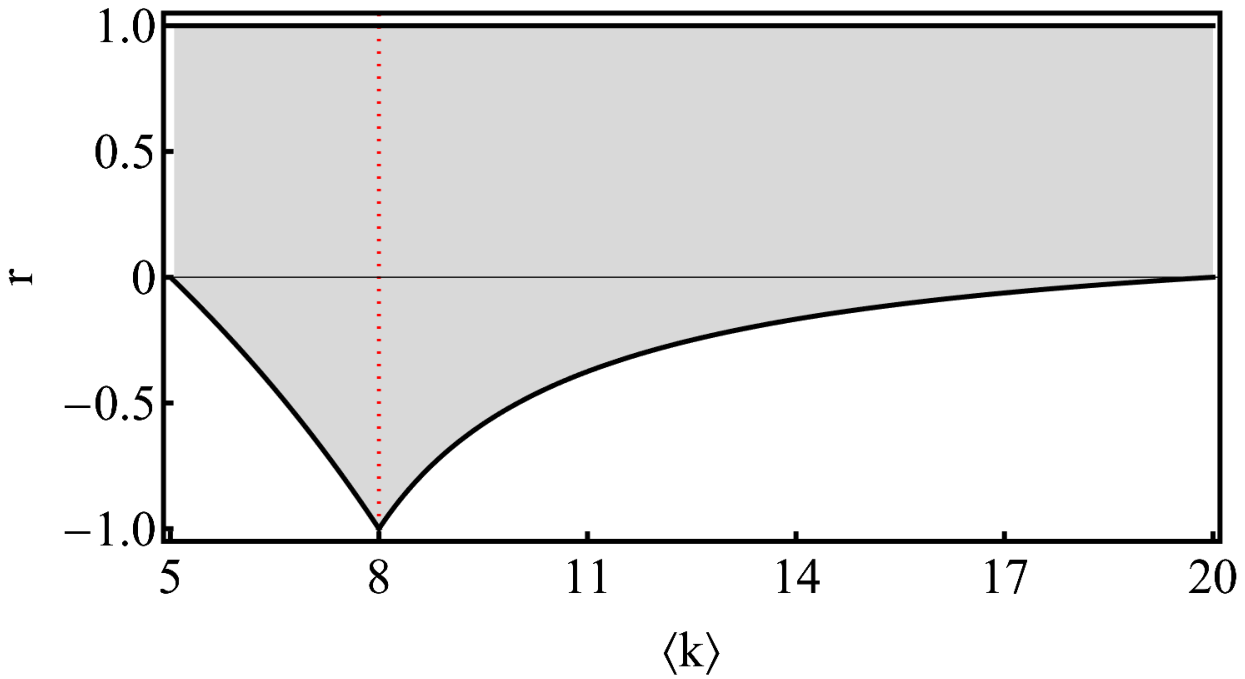}
(b)
\includegraphics[width=.60\textwidth]{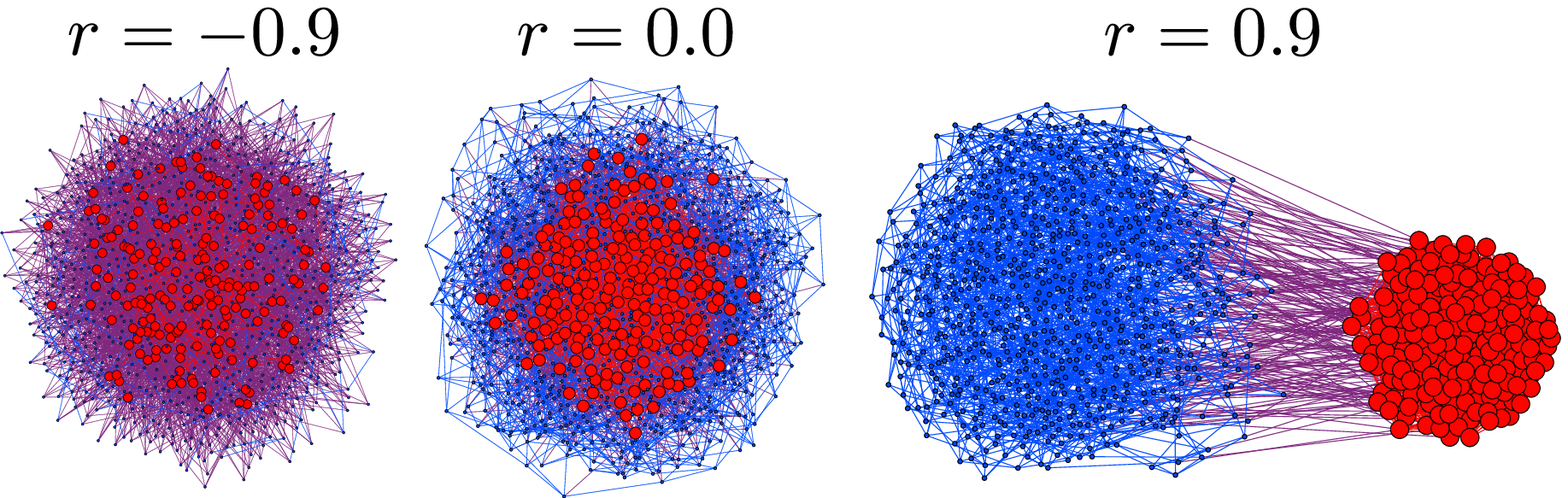}
\caption{
(a) Possible region (gray area) of the assortativity coefficient $r$ as a function of $\kav$ with $k_1=20$ and $k_2=5$. 
At $\kav=8$ (indicated by the red-dotted vertical line), the assortativity coefficient $r$ can range from $r=-1$ (perfectly disassortative) to $r=1$ (perfectly assortative).
(b) (Left to right): examples of a disassortative bimodal network ($r=-0.9$), a neutral bimodal network ($r=0.0$), and an assortative bimodal network ($r=0.9$).
All networks have $N=10^3$ nodes and $\kav=8$ (or equivalently $p_1=1-p_2=0.2$) with $k_1=20$ and $k_2=5$.
}
\label{fig:assortativityCoefficient}
\end{center}
\end{figure}

The assortativity coefficient $r$ \cite{newman2002assortative,newman2003mixing}, which quantifies the degree of assortative mixing in a network, is defined as
\be
r = \frac{\sum_{x} e_{xx} - \sum_{x} e_{xx}^{\rm rnd}}{1-\sum_{x} e_{xx}^{\rm rnd}}, \label{eq:newmanR}
\ee
where $e_{xy}$ is the probability that a randomly chosen edge in a given network  connects a node of type $x$ to one of type $y$, $e_{xy}=P_e(k_x,k_y)$, and $e_{xy}^{\rm rnd}=P_e(k_x)P_e(k_y)$ 
is the probability that a randomly chosen edge in the corresponding randomized network connects a node of type $x$ to one of type $y$.
When $r>0$, $r=0$, and $r<0$, the network is assortative, neutral, and disassortative, respectively. 
On correlated bimodal networks, Eq.~(\ref{eq:newmanR}) (after some transformations) simplifies to
\be
r=1-\frac{\langle k \rangle}{p_1 k_1}p_{21}. \label{eq:rBimodal}
\ee
Another assortativity coefficient $r_{kk'}$ is the Pearson's correlation coefficient of nearest degrees:
\be
r_{kk'} = \frac{4 \langle k k' \rangle_e - \langle k+k' \rangle_e^2}{2 \langle k^2+k'^2 \rangle_e - \langle k+k' \rangle_e^2} 
= \frac{\langle k k' \rangle_e - \langle k\rangle_e^2}{\langle k^2\rangle_e - \langle k\rangle_e^2},
\ee
where $\langle f(k,k')\rangle_e=\frac{1}{2} \sum_{k,k'} f(k,k')P_e(k,k')$. 
The Pearson's coefficient $r_{kk'}$ measures the degree-degree correlations, but we note that it gives the assortativity coefficient $r$ of a correlated bimodal network~\cite{mizutaka2016robustness}.

Figure~\ref{fig:assortativityCoefficient} (a) plots the assortativity coefficient $r$ as a function of $\langle k\rangle$ on bimodal networks with $k_{1}=20$ and $k_{2}=5$.
The possible region is the gray region in the plot. 
The maximum assortativity of a bimodal network is always $r_{\rm max}=1$.
All nodes connect with nodes of the same type, and there is no interconnectivity between the type-$1$ and type-$2$ nodes.  
The networks are then separated into two components, one comprising all type-$1$ nodes and the other consisting of all type-$2$ nodes. 
Meanwhile, the minimum assortativity of a bimodal network depends on the degree distribution~(\ref{eq:Pk}): $r_{\rm min}=-p_{1}k_{1}/p_{2}k_{2}$ for $p_{1}k_{1} < p_{2}k_{2}$, $r_{\rm min}=-p_{2}k_{2}/p_{1}k_{1}$ for $p_{1}k_{1}>p_{2}k_{2}$, and $r_{\rm min}=-1$ for $p_1 k_1=p_2 k_2$.
In the following section, we construct bimodal networks with $k_{1}=20$, $k_{2}=5$, and average degree $\langle k\rangle=8$.
On these networks, $p_1 k_1=p_2 k_2$ so the assortativity $r$ values lie within the range $[-1,1]$.

The bimodal networks with the desired assortativity $r$ is computationally generated as follows. 
Prepare $Np_{1}$ type-1 nodes and $Np_{2}$ type-2 nodes ($N$ nodes in total).
Assign $k_{1}$ stubs to each type-1 node and $k_{2}$ stubs to each type-2 node.
Form an edge by randomly selecting a stub from the type-1 nodes and a stub from the type-2 nodes and connecting them.
Repeat this process $Nk_{2}p_{2}p_{21}$ times, where $p_{21}$ is determined from Eq.~(\ref{eq:rBimodal}) with a given $r$
(note that $Nk_{2}p_{2}p_{21}=Nk_{1}p_{1}p_{12}$ edges exist between the type-1 and type-2 nodes).
Form further edges by randomly selecting two residual stubs of the type-1 and type-2 nodes and connecting them until no stubs remain.
Examples of bimodal networks with $r=-0.9,\;0.0,\;0.9$ are displayed in Fig.~\ref{fig:assortativityCoefficient} (b).
The strongly assortative network with $r=0.9$ is divided into two separate communities of type-1 nodes and type-2 nodes.
This case leads to double explosive spreadings in the number of infected nodes, as shown in the next section.

\section{Analysis}

This section investigates the effect of degree correlation on synergistic epidemics through the correlated bimodal networks.
To establish a qualitative picture, we first develop a mean-field approximation of the synergistic SIS on bimodal networks. 
In this approximation, we assume that the probability of a node being infected at time $t$ depends only on its node type $x \in \{1,2\}$.
Let $i_x(t)$ and $s_x(t) (=1-i_x(t))$ be the probabilities of a type-$x$ node being infected and susceptible, respectively, at time $t$.
Noting that a type-$x$ node connects to a type-$y$ node with probability $p_{xy}$ ($x,y \in \{1,2\}$), the probability $q_{x}dt$ of a susceptible type-$x$ node becoming infected within a small interval $dt$ is given by
\begin{align}
	q_{x} dt=\sum_{l=0}^{k_{x}}\binom{k_{x}}{l}p_{x1}^{l}p_{x2}^{k_{x}-l}\sum_{m=0}^{l}\binom{l}{m}i_{1}^{m}(1-i_{1})^{l-m}\sum_{n=0}^{k_{x}-l}\binom{k_{x}-l}{n}i_{2}^{n}(1-i_{2})^{k_{x}-l-n}(m+n)\lambda_{m+n} dt.
	\label{eq:qx}
\end{align}
Here $\lambda_{m+n}$ is the infection rate given by Eq.~(\ref{eq:infectionrate}), $l$ is a dummy variable indicating the number of type-$x$ neighbors, and $m$ and $n$ are dummy variables indicating the numbers of infected type-1 and type-2 neighbors, respectively.
Inserting Eq.~(\ref{eq:infectionrate}) into Eq.~(\ref{eq:qx}), we obtain the reduced form
\begin{align}
	q_{x}=\lambda k_{x}(p_{x1}i_{1}+p_{x2}i_{2})+\alpha\lambda k_{x}(k_{x}-1)(p_{x1}i_{1}+p_{x2}i_{2})^{2}.
	\label{eq:qx2}
\end{align}
From Eq.~(\ref{eq:qx2}), the time evolutions of infected densities $i_1(t)$ and $i_2(t)$ in the synergistic SIS model are respectively determined as
\begin{subequations}
\label{eq:MF-bimodal}
\begin{align}
\frac{d}{dt}i_1(t) 
&= -\mu i_1(t) +q_1 (1-i_1(t)) \nonumber \\
&= 
-\mu i_1 + (1 - i_1) (\lambda k_1 (p_{11} i_1+ p_{12} i_2) 
+ \alpha \lambda k_1 (k_1 - 1) (p_{11} i_1+ p_{12} i_2)^2), \\
\frac{d}{dt}i_2(t) 
&= -\mu i_2(t) +q_2 (1-i_2(t)) \nonumber \\
&= 
-\mu i_2 + (1 - i_2) (\lambda k_2 (p_{21} i_1+ p_{22} i_2) 
+ \alpha \lambda k_2 (k_2 - 1) (p_{21} i_1+ p_{22} i_2)^2).
\end{align}
\end{subequations}
Without loss of generality, we set $\mu=1$.
The total density $i(t)$ of the infected nodes at time $t$ is given as
\be
i(t)=p_1 i_1(t)+p_2 i_2(t).
\ee
The total density $s(t)=p_1 s_1(t)+p_2 s_2(t)$ of susceptible nodes at time $t$ satisfies the conservation law $s(t)+i(t)=1$.
The SIS process starts from a random initial condition in which each node is initially infected with probability $\rho$ and is susceptible otherwise, giving $i_1(0)=i_2(0)=i(0)=\rho$.
The system evolves and eventually converges to a steady state.
We are interested in the steady-state infected density $i(\infty)$ (the density of infected nodes $i(t)$ in the long time limit $t \to \infty$). 

\begin{figure}[t]
\begin{center}
(a)
\includegraphics[width=.3\textwidth]{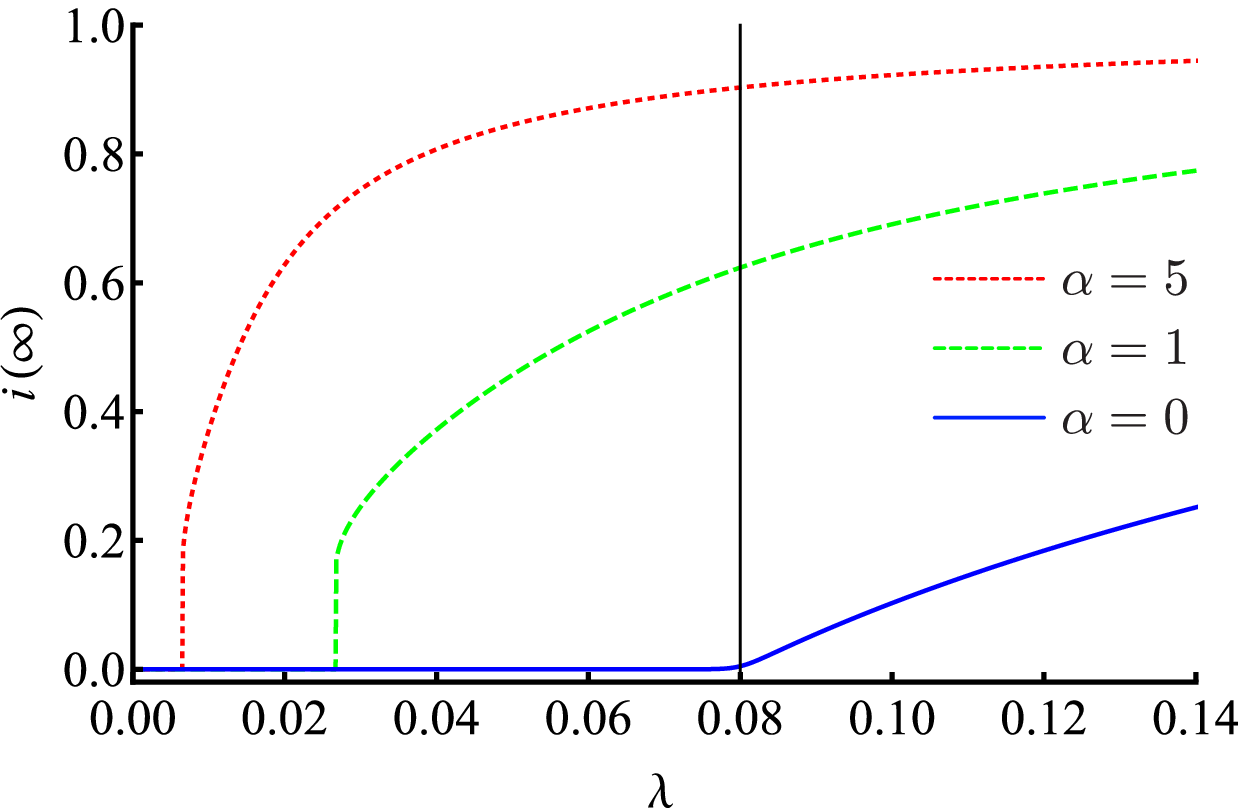}
(b)
\includegraphics[width=.3\textwidth]{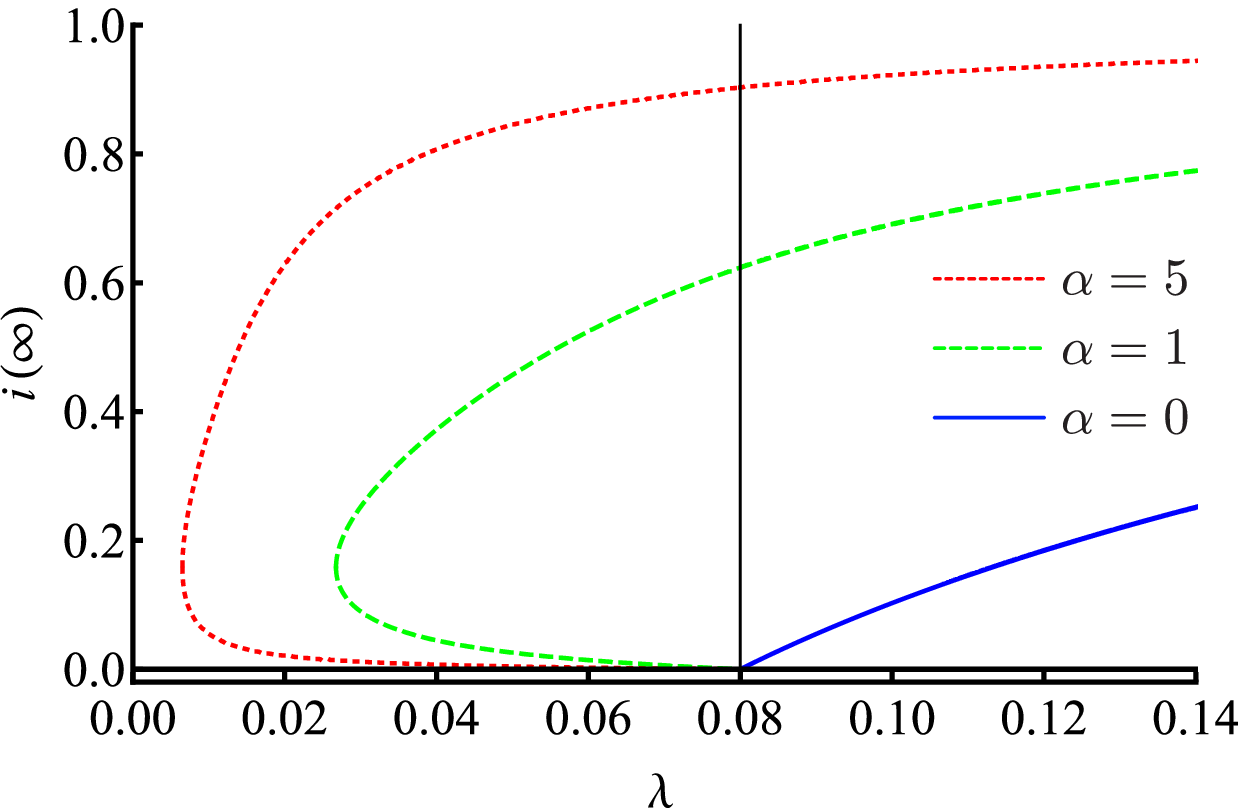}
(c)
\includegraphics[width=.30\textwidth]{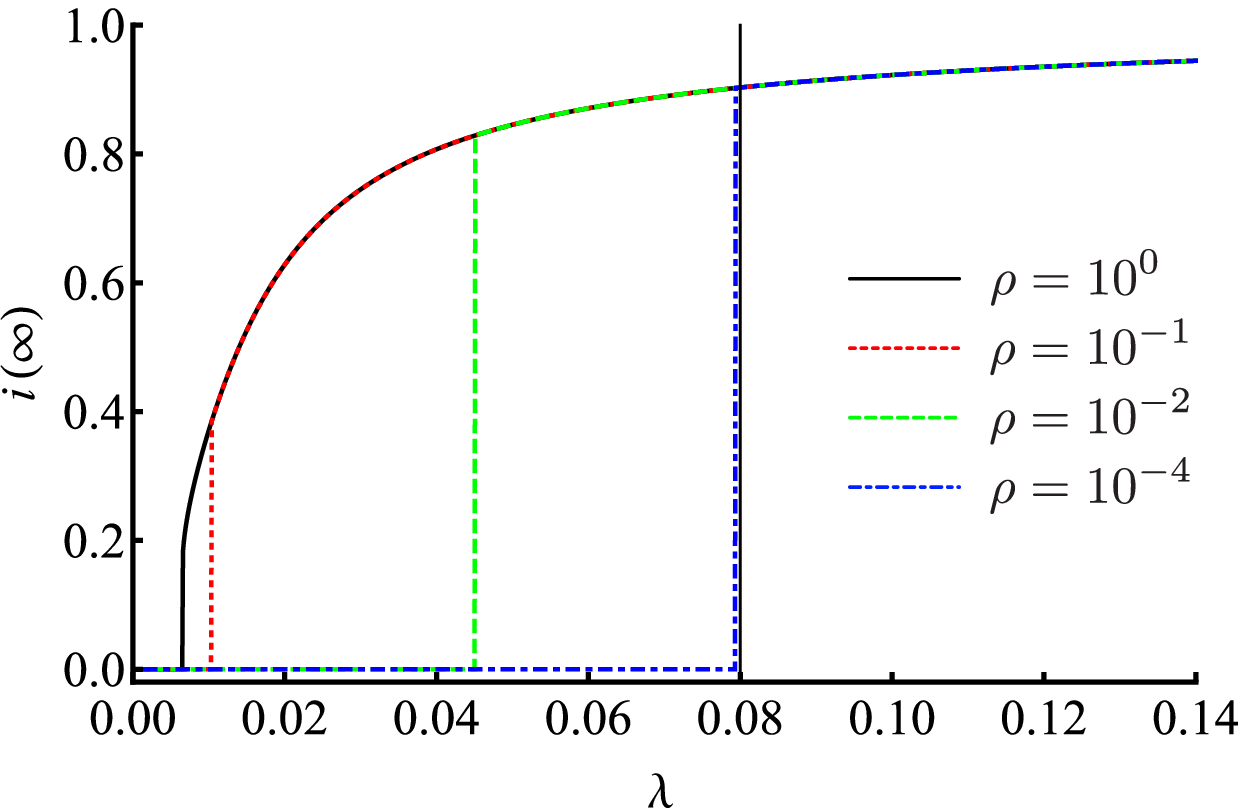}
\caption{
Mean-field results of the synergistic SIS model on the neutral bimodal network ($r=0$).
(a) Steady-state infected density $i(\infty)$ when all nodes are initially infected ($\rho=1$) 
and (b) possible infected densities given by the fixed points of Eq.~(\ref{eq:MF-bimodal}). 
In (a) and (b), the red-dotted, green-dashed, and blue-solid lines are the results of $\alpha = 5$, $\alpha = 1$, and $\alpha=0$, respectively.
(c) Steady-state infected density $i(\infty)$ of the synergistic SIS model with $\alpha=5$ under different initial conditions: $\rho=1$ (black-solid line), $\rho=10^{-1}$ (red-dotted line), $\rho=10^{-2}$ (green-dashed line), and $\rho=10^{-4}$ (blue-dotted-dashed line).
The vertical thin line is located at $\lambda_g = 0.08$ above which the system is in the endemic state.
}
\label{fig:infectedDensity:neutralBimodal}
\end{center}
\end{figure}

To confirm typical behaviors of the synergistic SIS model, we first simulate the model on the neutral bimodal network ($r=0$).
Figure~\ref{fig:infectedDensity:neutralBimodal} (a) plots the steady-state infected density $i(\infty)$ of the synergistic SIS model on the neutral bimodal network, determined by the mean-field equation when all nodes are initially infected ($\rho=1$).
In the presence of synergy ($\alpha=1, 5$), the present model exhibits an explosive spreading, i.e., $i(\infty)$ jumps discontinuously from $i(\infty)=0$ to $i(\infty) \neq 0$ at a certain $\lambda$.
On the other hand, in the absence of synergy ($\alpha=0$), $i(\infty)$ increases continuously from zero as $\lambda$ increases.

The possible values of $i(\infty)$ are obtained by solving Eq.~(\ref{eq:MF-bimodal}) with $d i_1 (t)/dt= d i_2 (t)/dt=0$.
Figure~\ref{fig:infectedDensity:neutralBimodal} (b) plots the possible infected densities when $\alpha=0$, $1$, and $5$.
The plot of the synergistic SIS model with $\alpha=5$ (red-dotted line) presents three regions.
In region I ($0<\lambda<\lambda_l \approx 0.0066$), a unique stable fixed point exists at $i(\infty)=0$, affirming that the infection becomes extinct.
In region II ($\lambda>\lambda_g =0.08$\footnote{From the condition that a trivial fixed point $(i^{*}_{1},i^{*}_{2})=(0,0)$ becomes unstable for $\lambda>\lambda_{g}$, $\lambda_{g}$ is a solution of $rk_1 k_2 \lambda^2-(p_{11}k_1 +p_{22}k_2)\lambda +1=0$.}), an unstable fixed point and a stable fixed point are found at $i(\infty)=0$ and $i(\infty)>0$, respectively, meaning that the system reaches an endemic state.
In region III ($\lambda_l<\lambda<\lambda_g$), Eq.~(\ref{eq:MF-bimodal}) generates three fixed points, two stable points located at $i(\infty)=0$ and $i(\infty)>0$ and one unstable point located between the stable fixed points.
Within this bistable region, hysteresis emerges \cite{gomez2016explosive} and the system reaches an extinct or endemic state depending on the initial $\rho$.
In the model with synergy (Fig.~\ref{fig:infectedDensity:neutralBimodal} (c)), $i(\infty)$ discontinuously jumps at some $\rho$-dependent $\lambda$.
Explosive spreading and hysteresis are also observed at $\alpha=1$ (green-dashed lines in Figs.~\ref{fig:infectedDensity:neutralBimodal} (a) and (b)).
In contrast, $i(\infty)$ in the classical SIS model ($\alpha=0$; see blue-solid lines in Figs.~\ref{fig:infectedDensity:neutralBimodal} (a) and (b)) shows no discontinuity and no bistability.

\begin{figure}[t]
\begin{center}
(a)
\includegraphics[width=.30\textwidth]{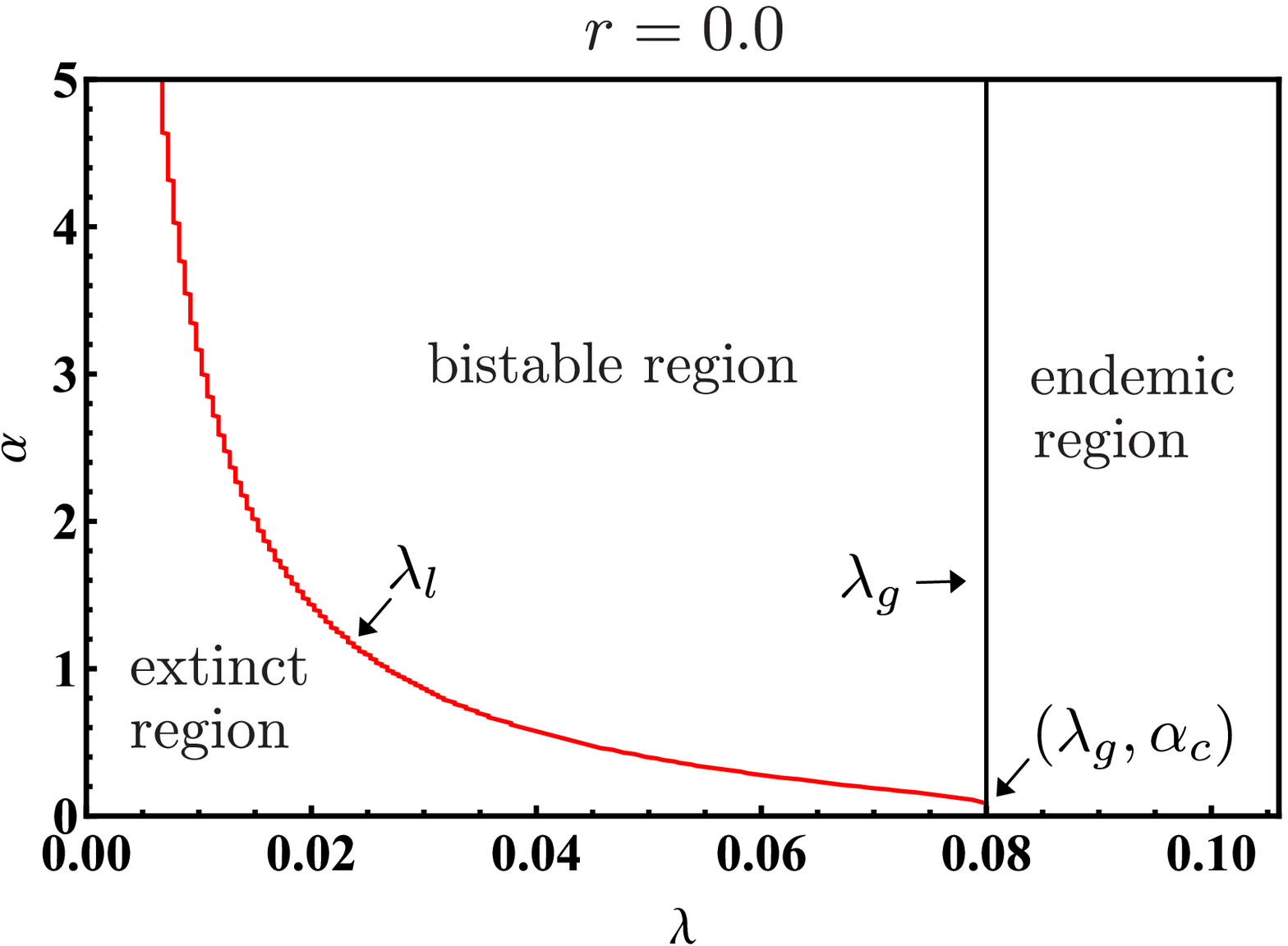}
(b)
\includegraphics[width=.30\textwidth]{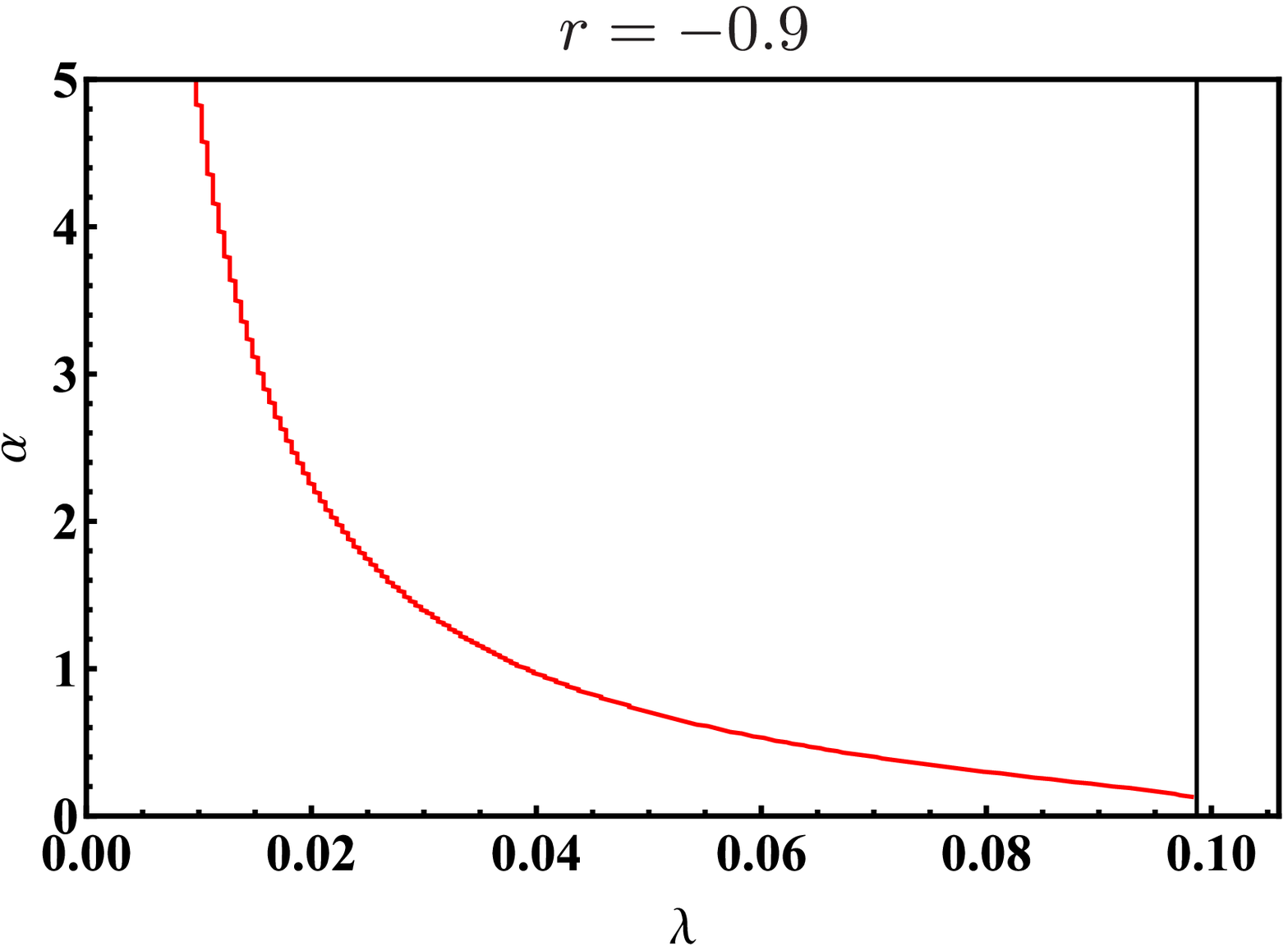}
(c)
\includegraphics[width=.30\textwidth]{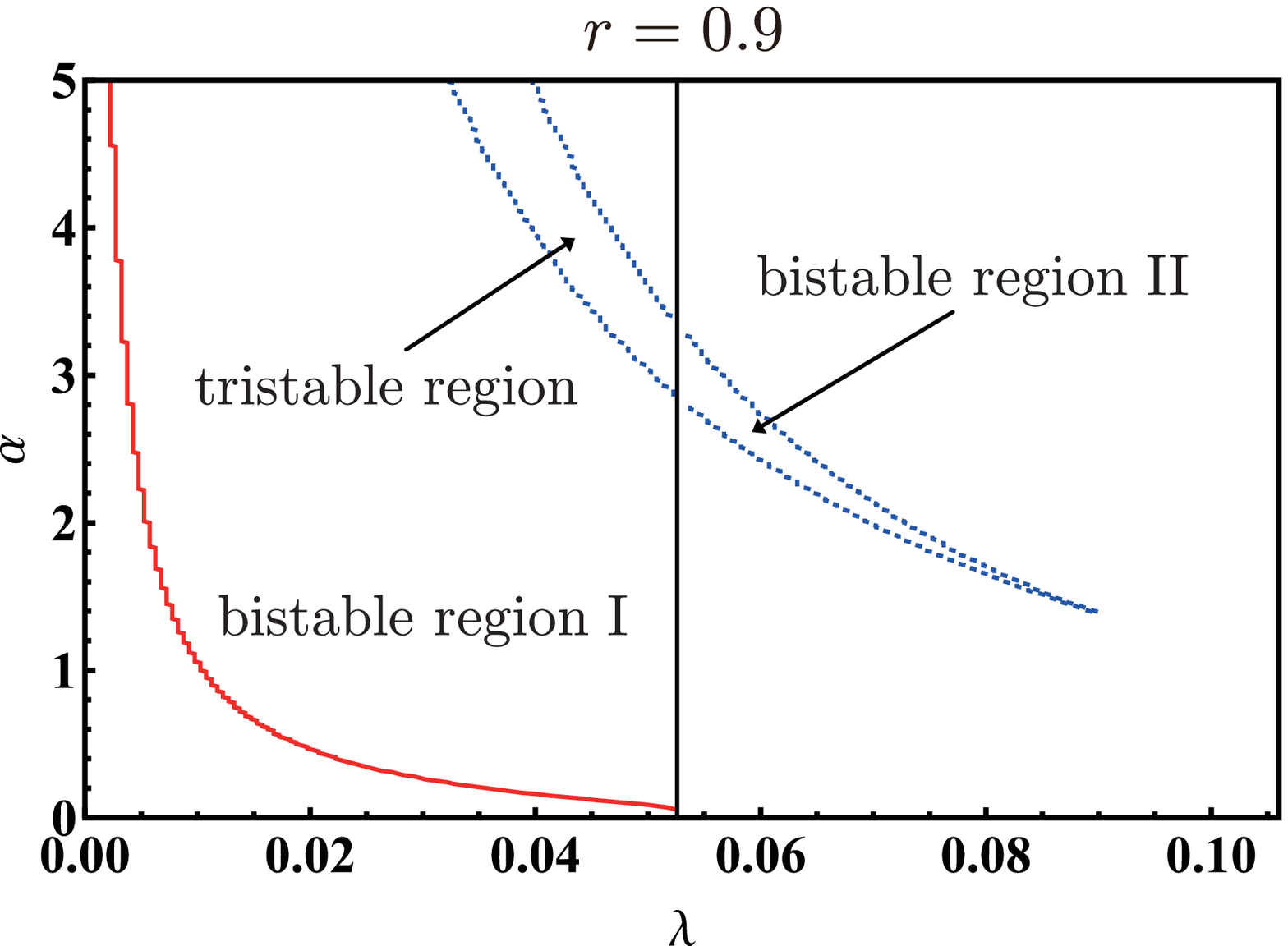}
\caption{
Phase diagrams in the $(\lambda,\alpha)$ plane for the synergistic SIS model on the correlated bimodal networks with (a) $r=0.0$, (b) $r=-0.9$, and (c) $r=0.9$. 
The lines are for different numbers of fixed points of Eq.~(\ref{eq:MF-bimodal}).
For the extinct region, infected density reaches a trivial stable fixed point, $\rho(\infty) \to 0$, irrespective of the value of the initial infected density $\rho >0$. 
The infection becomes extinct or persists in the bistable region depending on $\rho$, which means two stable fixed points exist. 
In the endemic region, the infection persists for $\rho>0$.
In the bistable region I, the infection becomes persistent in $k_1$-community (see Fig.~\ref{fig:assortativityCoefficient}(b)) or extinct depending on the initial condition. 
Three stable fixed points emerge in the tristable region. The behaviors are more complex than in other regions (see Figs.~\ref{fig:infectedDensity:assortativeBimodal} (b) and \ref{fig:bifurcationDiagram-fixedAlpha}(c)). 
In the bistable region II, the infection persists in $k_2$-community depending on the value of $\rho$, while it always persists in $k_1$-community for $\rho >0$.
}
\label{fig:phaseBoundarySeveralR}
\end{center}
\end{figure}

\begin{figure}[t]
\begin{center}
\includegraphics[width=.45\textwidth]{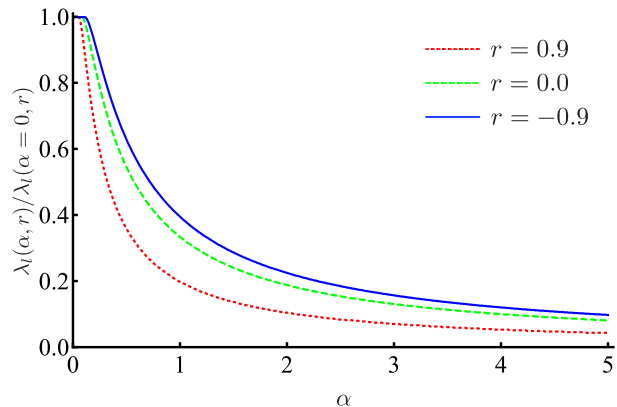}
\caption{
$\alpha$ dependence of threshold $\lambda_{l}(\alpha,r)$ divided by $\lambda_{l}(\alpha=0,r)$, which is the epidemic threshold of the classical SIS model. The red-dotted, green-dashed, and blue-solid lines plot the results of $r=-0.9$, $r = 0$, and $r=0.9$, respectively. The results are obtained by the mean-filed treatment.
}
\label{fig:Comparison}
\end{center}
\end{figure}

By counting the number of fixed points of Eq.~(\ref{eq:MF-bimodal}), we determine the phase diagrams of the synergistic SIS model on the correlated bimodal networks (Figs.~\ref{fig:phaseBoundarySeveralR} (a)--(c)).
The neutral bimodal network has a phase diagram of two boundaries $\lambda_l$ and $\lambda_g$ that separate extinct, bistable, and endemic regions (Fig.~\ref{fig:phaseBoundarySeveralR} (a)).
As $\alpha$ increases from a certain value $\alpha_c \approx 0.086$ in the mean-field approximation, $\lambda_l$ decreases but $\lambda_g$ remains constant ($\lambda_g=0.08$), informing that the bistable region broadens with increasing synergistic strength $\alpha$.
When $\alpha<\alpha_c$, the bistable region disappears, and one boundary $\lambda_l=\lambda_g$ separates the extinct and endemic regions.

Let us investigate how the phase diagram in the $(\lambda, \alpha)$-plane changes when a network has a degree-correlated structure.
Figures~\ref{fig:phaseBoundarySeveralR} (b) and (c) are phase diagrams of the strongly disassortative ($r=-0.9$) and strongly assortative ($r=0.9$) bimodal networks, respectively.
In the disassortative case, the synergistic SIS model behaves qualitatively the same as in the neutral case (with three regions separated by two boundaries $\lambda_l$ and $\lambda_g$) although the boundaries shift from the neutral case.
Specifically, $\lambda_l$ and $\lambda_g$ at any fixed $\alpha$ are larger in the disassortative case than in the neutral case. 
In contrast, the phase diagram of the assortative case qualitatively differs from the other cases and presents two additional boundaries (dotted lines in Fig.~\ref{fig:phaseBoundarySeveralR} (c)).
Figure~\ref{fig:Comparison} shows the $\alpha$ dependence of threshold $\lambda_{l}(\alpha,r)$ divided by $\lambda_{l}(\alpha=0,r)$, which is the epidemic threshold of the classical SIS model. The red-dotted, green-dashed, and blue-solid lines plot the results of $r=-0.9$, $r = 0$, and $r=0.9$, respectively. As seen from Fig.~\ref{fig:Comparison}, in each value of $\alpha$, the red-dotted (blue-solid) line is the lowest (highest) in the three lines, which means that the synergy acts better in networks with positive degree correlations than ones with negative degree correlations.

\begin{figure}[t]
\begin{center}
(a)
\includegraphics[width=.30\textwidth]{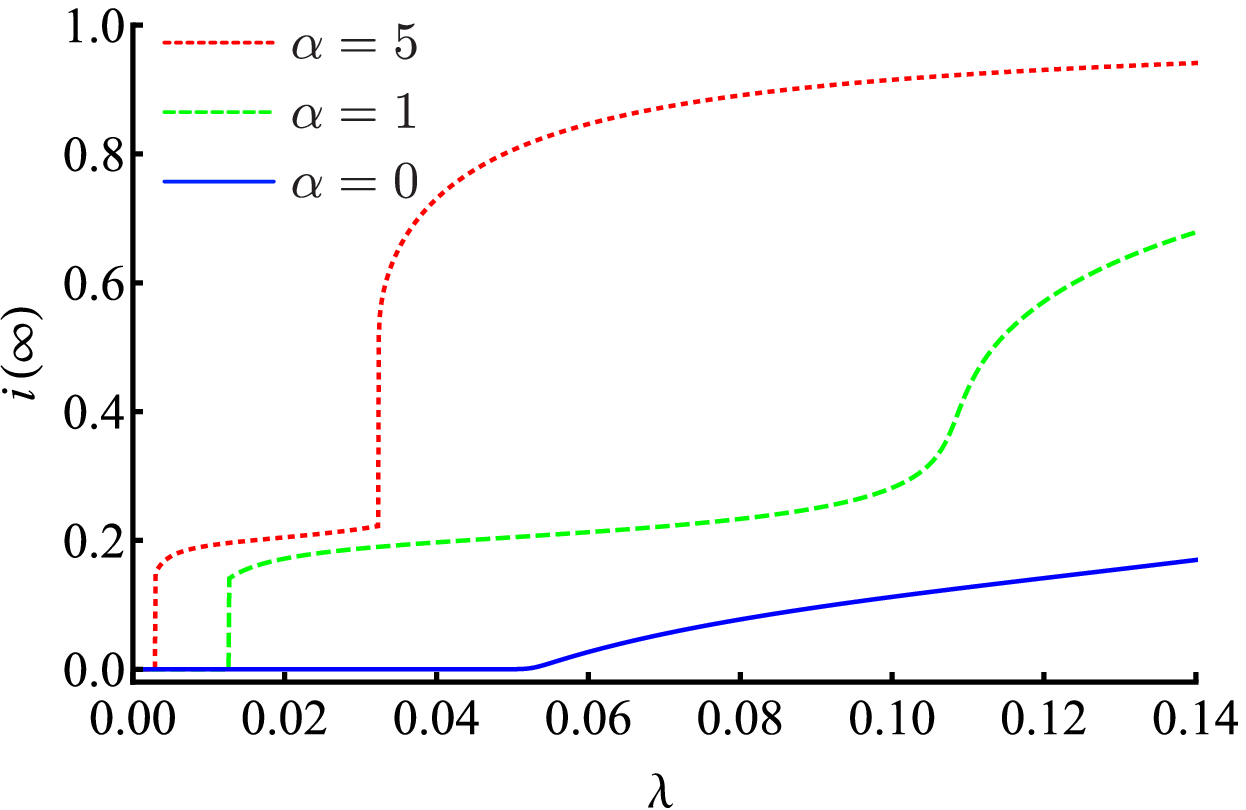}
(b)
\includegraphics[width=.30\textwidth]{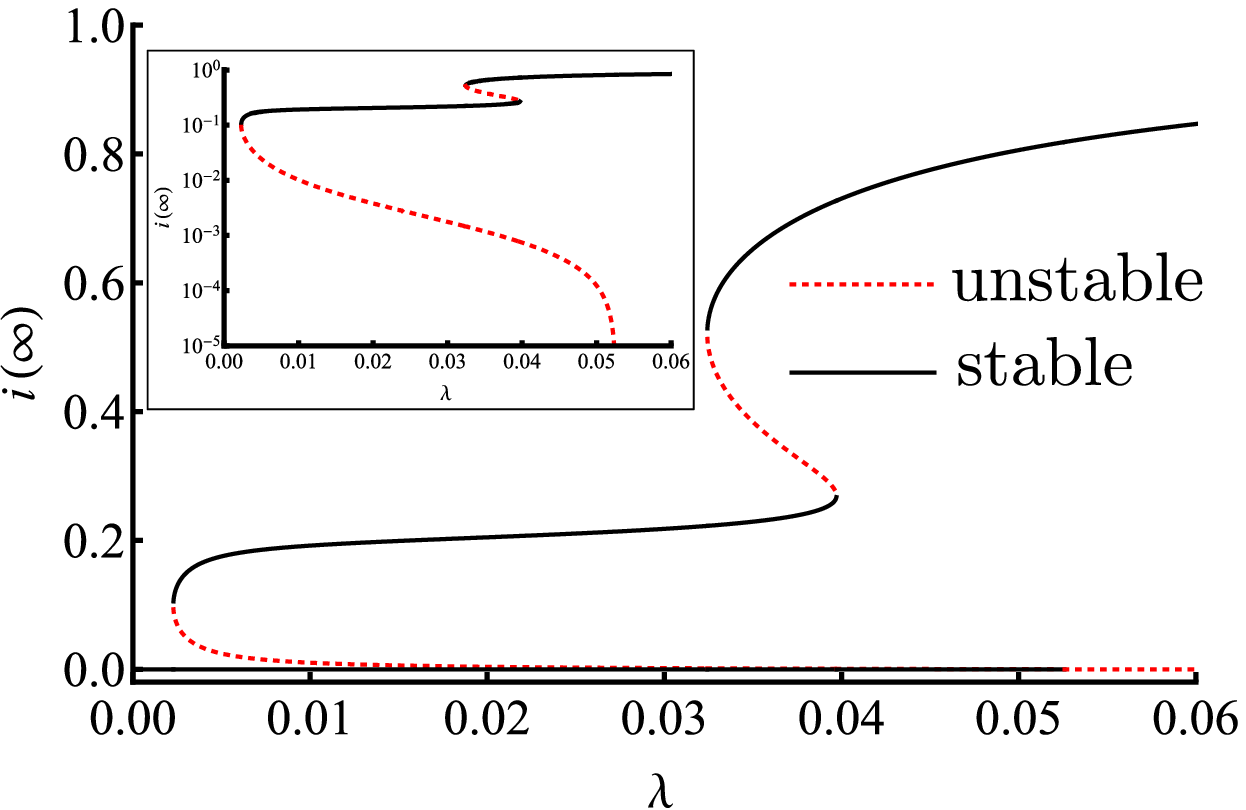}
(c)
\includegraphics[width=.30\textwidth]{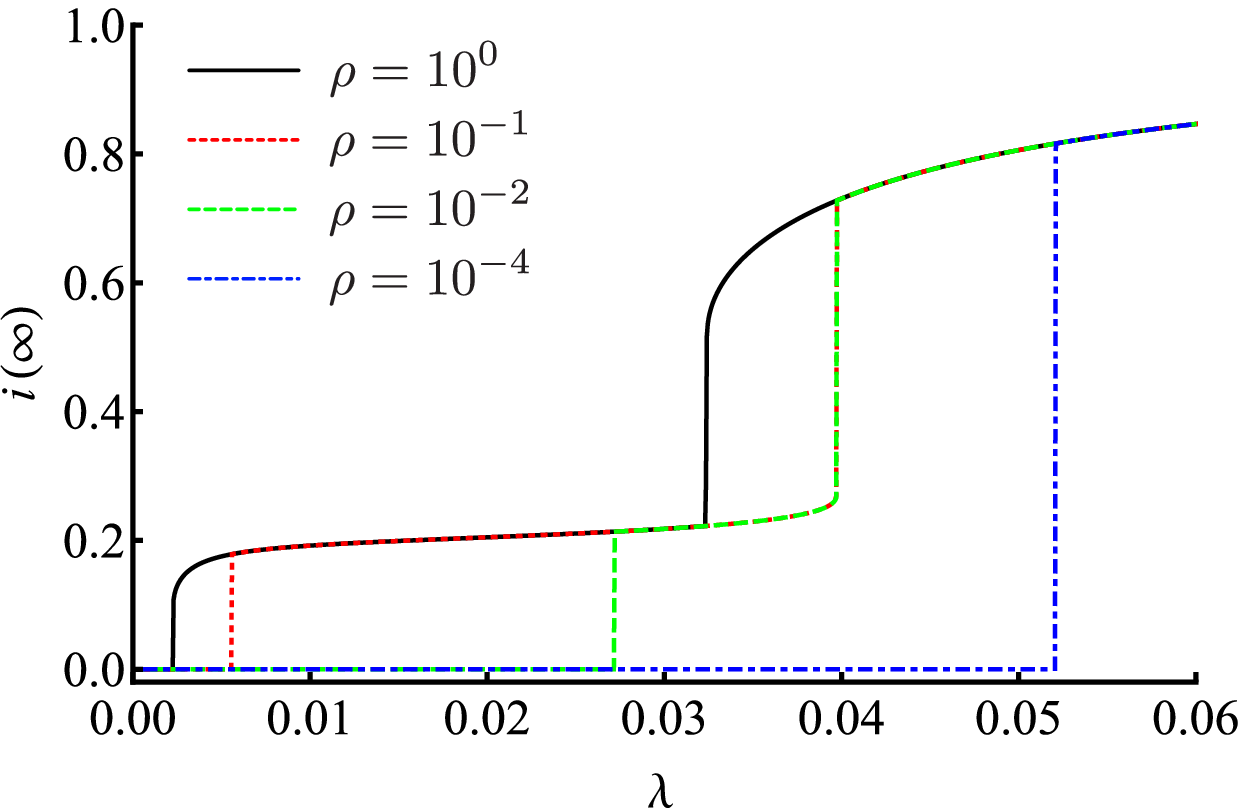}
\caption{
Mean-field results of the synergistic SIS model on the strongly assortative bimodal network ($r=0.9$).
(a) Steady-state infected density $i(\infty)$ when all nodes are initially infected ($\rho=1$). 
The red-dotted, green-dashed, and blue-solid lines plot the results of $\alpha = 5$, $\alpha = 1$, and $\alpha=0$, respectively.
(b) Possible infected densities determined from the fixed points of Eq.~(\ref{eq:MF-bimodal}) with $\alpha=5$.
The black-solid and red-dotted lines indicate stable and unstable fixed points, respectively.
(c) Steady-state infected density $i(\infty)$ with $\alpha=5$ under different initial conditions: $\rho=1$ (black-solid line), $\rho=10^{-1}$ (red-dotted line), $\rho=10^{-2}$ (green-dashed line), and $\rho=10^{-4}$ (blue-dotted-dashed line).
}
\label{fig:infectedDensity:assortativeBimodal}
\end{center}
\end{figure}

\begin{figure}[t]
\begin{center}
(a)
\includegraphics[width=.28\textwidth]{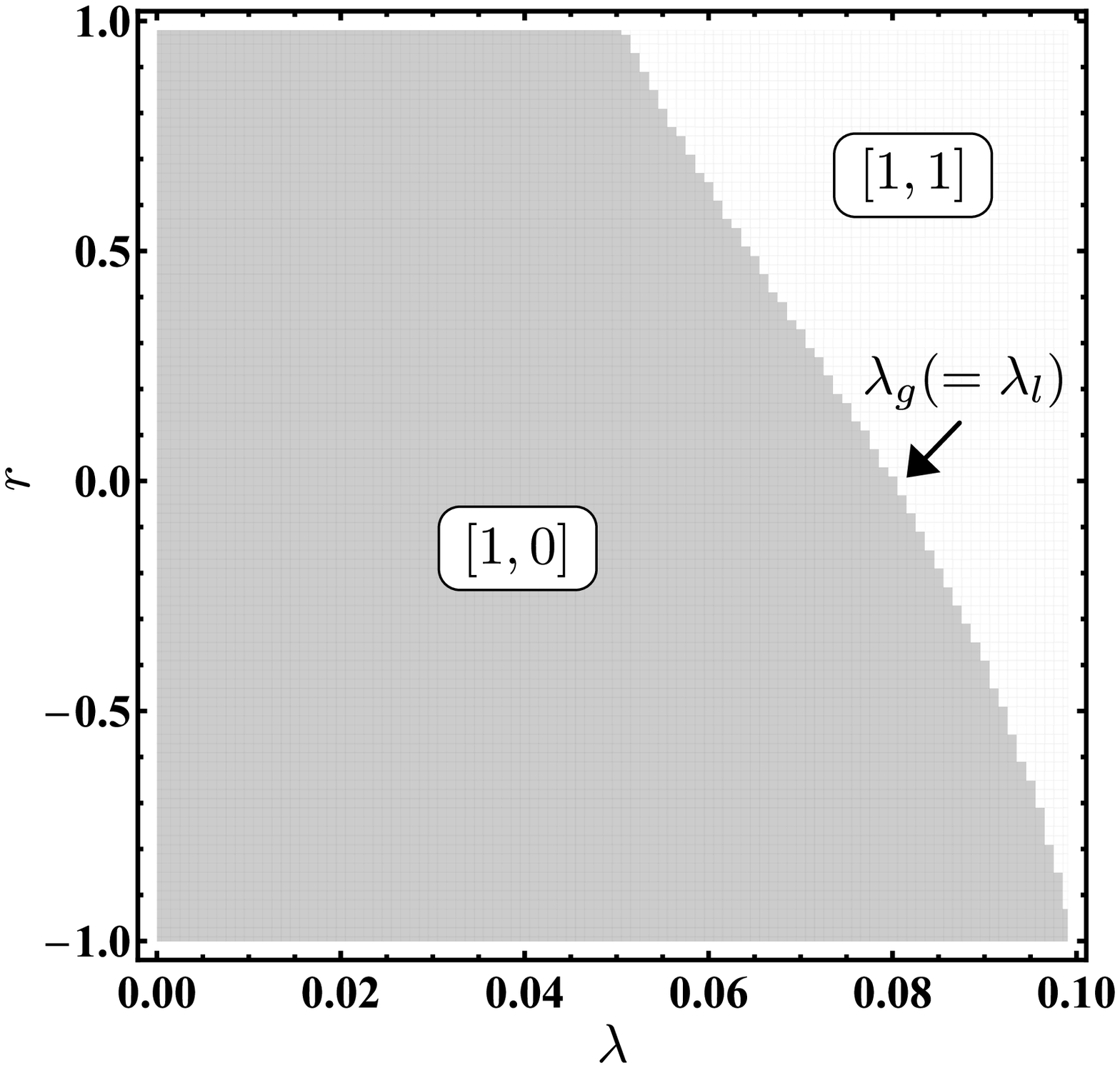}
(b)
\includegraphics[width=.28\textwidth]{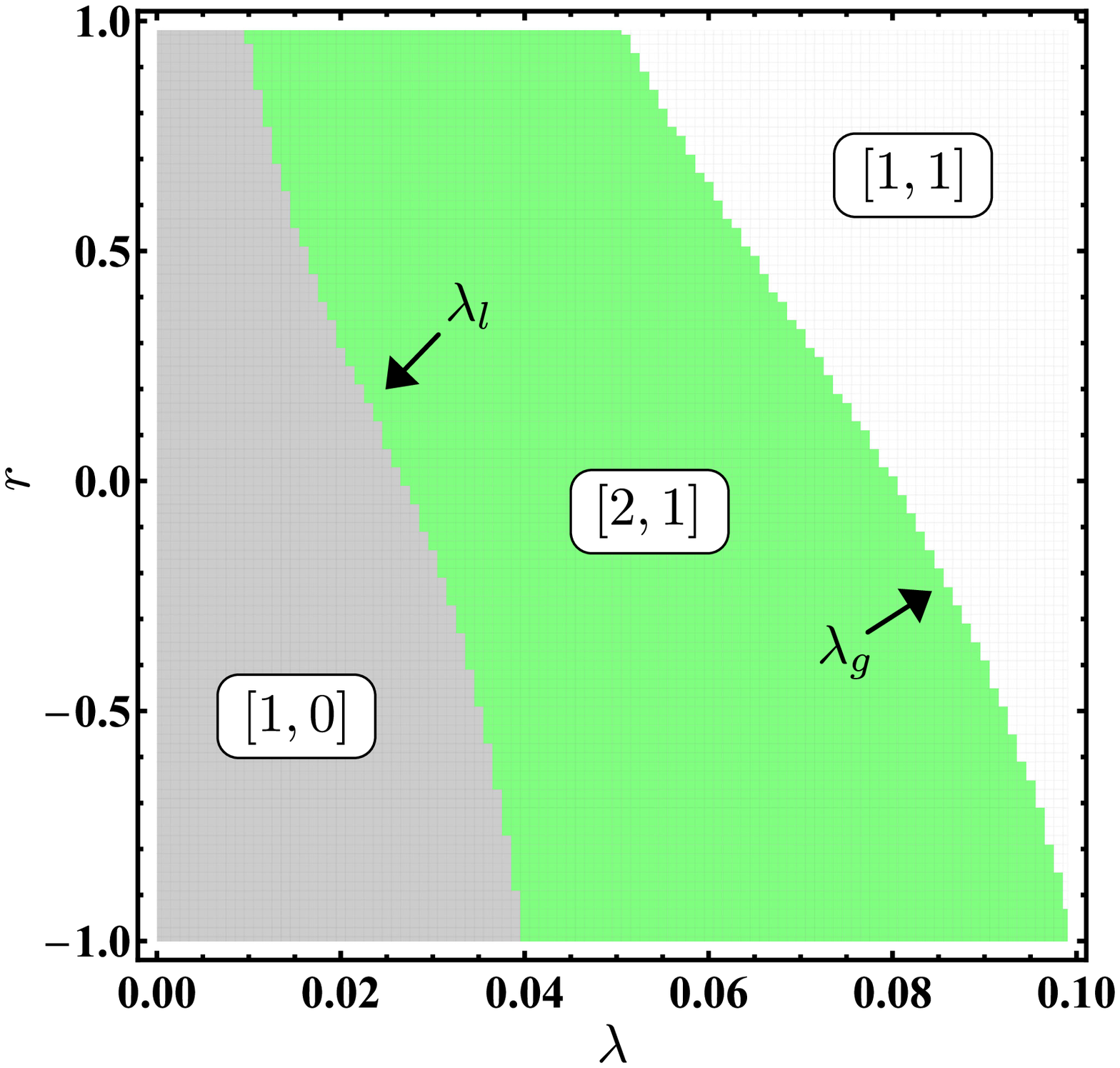}
(c)
\includegraphics[width=.28\textwidth]{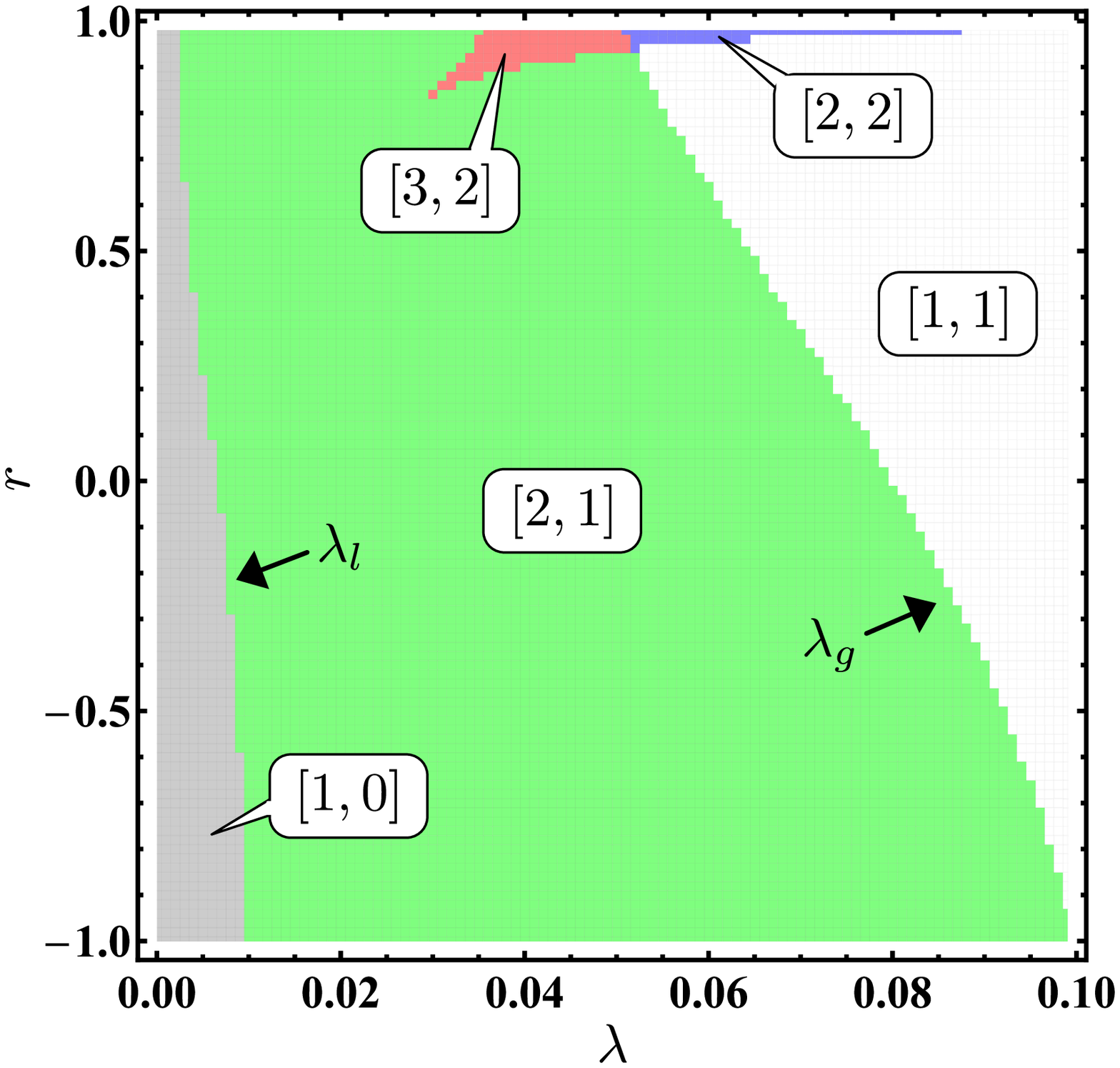}
\caption{
Phase diagrams in the $(\lambda,r)$ plane for (a) $\alpha=0$, (b) $\alpha=1$, and (c) $\alpha=5$.
The regions are divided in terms of the numbers of stable and unstable fixed points of Eq.~(\ref{eq:MF-bimodal}) (indicated by the first and second numbers within the square brackets, respectively).
}
\label{fig:bifurcationDiagram-fixedAlpha}
\end{center}
\end{figure}

Figure~\ref{fig:infectedDensity:assortativeBimodal} shows the steady-state infected density of the synergistic SIS model on the strongly assortative bimodal network. 
As the blue-solid and green-dashed lines in Fig.~\ref{fig:infectedDensity:assortativeBimodal} (a) show, an ordinary epidemic spreading and an explosive spreading are observed even for a strongly assortative bimodal network when synergy is absent ($\alpha=0$) and is weakly present ($\alpha=1$), respectively.
The spreading behavior changes under strong synergy ($\alpha=5$): the present model exhibits a discontinuous jump in $i(\infty)$ at small $\lambda$ and another discontinuous jump of $i(\infty)$ at larger $\lambda$ (see the red-dotted line in Fig.~\ref{fig:infectedDensity:assortativeBimodal} (a)).
These two explosive spreadings are actually discontinuous, as evidenced by the fixed points of Eq.~(\ref{eq:MF-bimodal}) (Fig.~\ref{fig:infectedDensity:assortativeBimodal}(b)).

The two jumps, in the case of $\alpha=5$, are attributable to the splitting of the strongly assortative bimodal network into two communities (a community of type-1 nodes with a large degree and a community of of type-2 nodes with a small degree).
When starting with $\rho=1$, the explosive spreading occurs and the infection persists in the community of type-1 nodes at $\lambda \approx 0.003$.
However, the infection rate for the first explosive spreading is still too small for the infection to persist in the community of type-2 nodes.
At a larger value of $\lambda$ ($\lambda \approx 0.032$), the community of type-2 nodes undergoes another explosive spreading, and the infection becomes endemic in that community.
Also, whether explosive spreading is triggered in each community depended on the initial seed fraction. 
Consequently, the system develops complex hysteresis behavior (Fig.~\ref{fig:infectedDensity:assortativeBimodal}(c)).

Let us now explore the effect of degree correlation on transitions in the synergistic SIS model. 
Panels (a), (b), and (c) in Fig.~\ref{fig:bifurcationDiagram-fixedAlpha} show the phase diagrams in the $(\lambda, r)$ plane for the models with $\alpha=0$ (without synergy), $\alpha=1$ (weak synergy), and $\alpha=5$ (strong synergy), respectively.
In panel (a), the boundary separating the extinct (gray-colored) and endemic (white-colored) regions is the well-known epidemic threshold \cite{pastor2001epidemic,castellano2010thresholds}.
The epidemic threshold is a decreasing function of the assortativity $r$.
Infectious diseases invade and persist more easily on assortative bimodal networks than on disassortative ones.
This tendency has been previously reported \cite{van2010influence,morita2021solvable}.
In Fig.~\ref{fig:bifurcationDiagram-fixedAlpha} (b), the bistable (green-colored) region appearing between the extinct and endemic regions is attributable to the synergistic effect.
The extinct-bistable and bistable-endemic boundaries correspond to $\lambda_l$ and $\lambda_g$, respectively, in Fig.~\ref{fig:phaseBoundarySeveralR}.
Here $\lambda_{l}$ is the threshold above which infectious diseases persist when all nodes are initially infected, and die out when an infinitesimal fraction of the nodes are initially infected.
The value of this threshold depends on the strength $\alpha$ of the synergy.
Meanwhile, $\lambda_{g}$ is the threshold above which infectious diseases invade and persist when an infinitesimal fraction of nodes are infected.
This threshold is independent of $\alpha$.
As occurs in the classical SIS model, assortativity reduces the positions of both boundaries ($\lambda_l$ and $\lambda_g$).
As shown in Fig.~\ref{fig:bifurcationDiagram-fixedAlpha} (c), the SIS model with strong synergy behaves identically to that of weak synergy except in the case of strong positive degree correlation.
Two discontinuous jumps appear only when a strong synergistic effect works on a strongly assortative networks ($r \gtrsim 0.8$).
The hysteresis region associated with the second explosive spreading (red and blue regions in Fig.~\ref{fig:bifurcationDiagram-fixedAlpha} (c)) shifts toward a larger $\lambda$ as $r$ increases.
Recall that the second explosive spreading through the strongly assortative network occurs in the community of type-2 nodes with small degree $k_2 (< k_1)$.
As $r$ increases, the communities of type-2 and type-1 nodes (in which epidemics are already prevalent) becomes more weakly connected, and the number of infectious transmissions carried from type-1 nodes decreases. 
Therefore, as $r$ increases, a larger infection rate is required for the infection to penetrate or persist in the community of type-2 nodes.

Finally, we employ the Monte-Carlo simulations and approximate master equations (AMEs) \cite{lindquist2011effective,gleeson2013binary,hasegawa2016outbreaks} in order to confirm the predicted behaviors from the above-mentioned mean-field treatment quantitatively since the mean-field treatment is a rough approximation.
Gleeson \cite{gleeson2013binary} formulated AMEs of the classical SIS model on a network.
The AME approach accurately describes the synergistic SIS dynamics.
For our purpose, we consider the synergistic SIS model in infinitely large correlated bimodal networks. 
We denote by $s_{x,m}(t)$ and $i_{x,m}(t)$ the fractions of susceptible and infected nodes of type $x\in \{1,2\}$, respectively, having $m$ infected neighbors (by implication, $k_x-m$ susceptible neighbors) at time $t$. 
Similarly to \cite{gleeson2013binary}, we obtain the master equations for the evolution of each state density in the synergistic SIS model on the correlated bimodal network.
The equations are given by
\begin{subequations}
\label{eq:AME-bimodal}
\begin{align}
\frac{d}{dt} s_{1,m} &= -\lambda_m s_{1,m}+\mu i_{1,m} -\bssiA (k_1-m) s_{1,m} +\bssiA (k_1-m+1) s_{1,m-1} 
-\mu m s_{1,m} +\mu (m+1) s_{1,m+1},
\\
\frac{d}{dt} s_{2,m} &= -\lambda_m s_{2,m}+\mu i_{2,m} -\bssiB (k_2-m) s_{2,m} +\bssiB (k_2-m+1) s_{2,m-1} 
-\mu m s_{2,m} +\mu (m+1) s_{2,m+1},
\\
\frac{d}{dt} i_{1,m} &= -\mu i_{1,m} + \lambda_m s_{1,m} -\bisiA (k_1-m) i_{1,m} + \bisiA (k_1-m+1) i_{1,m-1}
-\mu m i_{1,m}+\mu (m+1) i_{1,m+1},
\\
\frac{d}{dt} i_{2,m} &= -\mu i_{2,m} + \lambda_m s_{2,m} -\bisiB (k_2-m) i_{2,m} + \bisiB (k_2-m+1) i_{2,m-1}
-\mu m i_{2,m}+\mu (m+1) i_{2,m+1}.
\end{align}
\end{subequations}
Here $\beta_{x, {\rm X}}^{\rm SI}$ represents the rate at which a randomly chosen susceptible neighbor of a randomly chosen type-$x$ node in state X (X $=$ S, I) switches its state from S to I.
As the present network has assortative mixing, 
the probability of a susceptible neighbor of a type-$x$ node being type $y$ is $p_{xy} s_y/ \sum_{y' \in \{1,2\}} p_{xy'} s_{y'}$, where $s_x$ is the probability of a type-$x$ node being susceptible at time $t$.
These time-dependent coefficients are as follows:
\be
\label{eq:coefficient-AME-bimodal}
\beta_{x,{\rm S}}^{\rm SI} =\frac{1}{\sum_{y' \in \{1,2\}} p_{xy'} s_{y'}} \sum_{y \in \{1,2\}} p_{xy} s_y \beta_y^{\rm SS}
\quad {\rm and} \quad
\beta_{x,{\rm I}}^{\rm SI} =\frac{1}{\sum_{y' \in \{1,2\}} p_{xy'} s_{y'}} \sum_{y \in \{1,2\}} p_{xy} s_y \beta_y^{\rm SI},
\ee
where $\beta_{y}^{\rm SS}$ ($\beta_{y}^{\rm SI}$) can be approximated as the rate at which an S-S (S-I) edge whose one end is type $y$ is changed to an S-I (I-I) edge as
\be
\label{eq:coefficient2-AME-bimodal}
\beta_y^{\rm SS}=\frac{\sum_{m=0}^{k_y} \lambda_m m (k_y-m) s_{y,m}}{\sum_{m=0}^{k_y} (k_{y}-m) s_{y,m}}
\quad {\rm and} \quad
\beta_y^{\rm SI}=\frac{\sum_{m=0}^{k_y} \lambda_m m^2 s_{y,m}}{\sum_{m=0}^{k_y} m s_{y,m}}.
\ee
Evaluating Eqs.~(\ref{eq:AME-bimodal}) in the initial state with seed fraction $\rho$, given by
\be
s_{x,m}(0)=(1-\rho) \binom{k_x}{m} \rho^m (1-\rho)^{k_x-m}
\quad {\rm and} \quad
i_{x,m}(0)=\rho \binom{k_x}{m} \rho^m (1-\rho)^{k_x-m},
\ee
for $x \in \{ 1,2\}$, the susceptible and infected densities of each node type at time $t$ are respectively given by
\be
s_1(t)=\sum_{m=0}^{k_1} s_{1,m}(t), 
\quad
i_1(t)=\sum_{m=0}^{k_1} i_{1,m}(t), 
\quad
s_2(t)=\sum_{m=0}^{k_2} s_{2,m}(t), 
\quad
i_2(t)=\sum_{m=0}^{k_2} i_{2,m}(t).
\ee

\begin{figure}[t]
\begin{center}
(a)
\includegraphics[width=.30\textwidth]{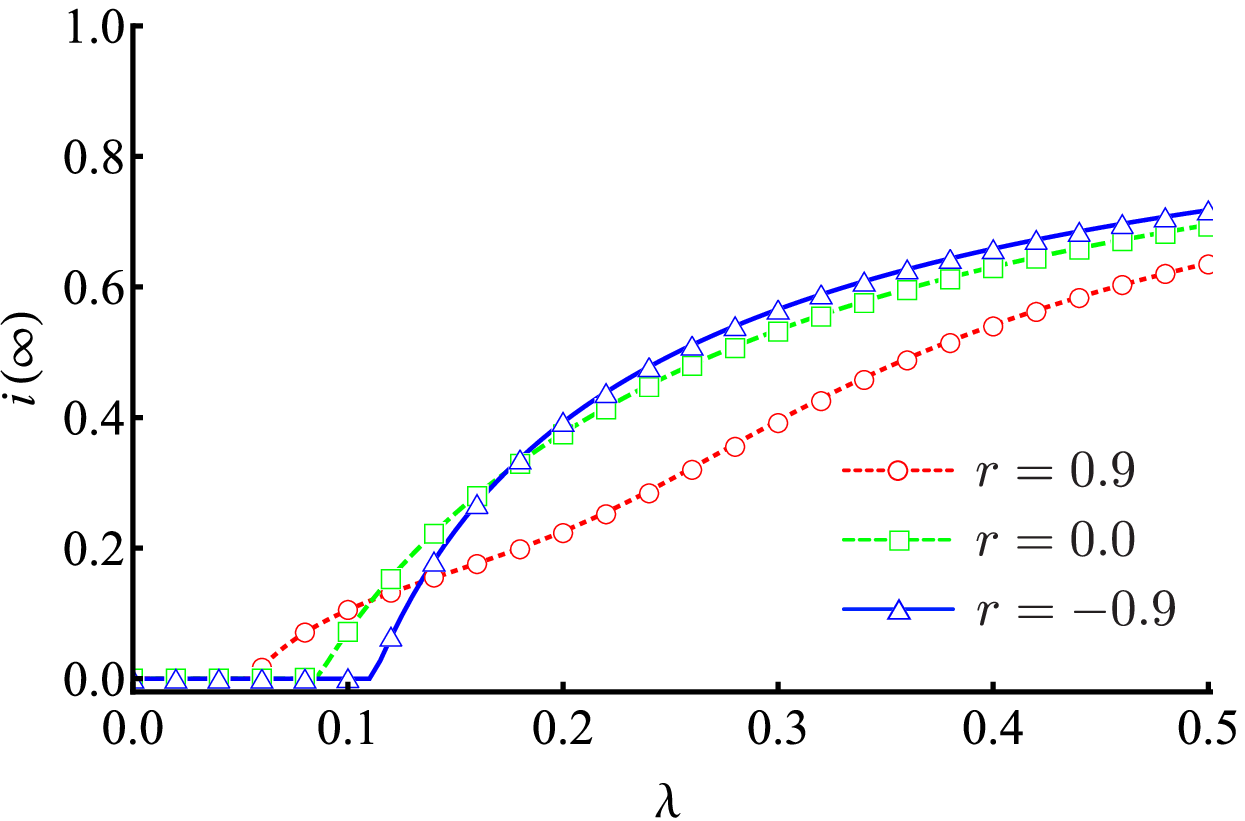}
(b)
\includegraphics[width=.30\textwidth]{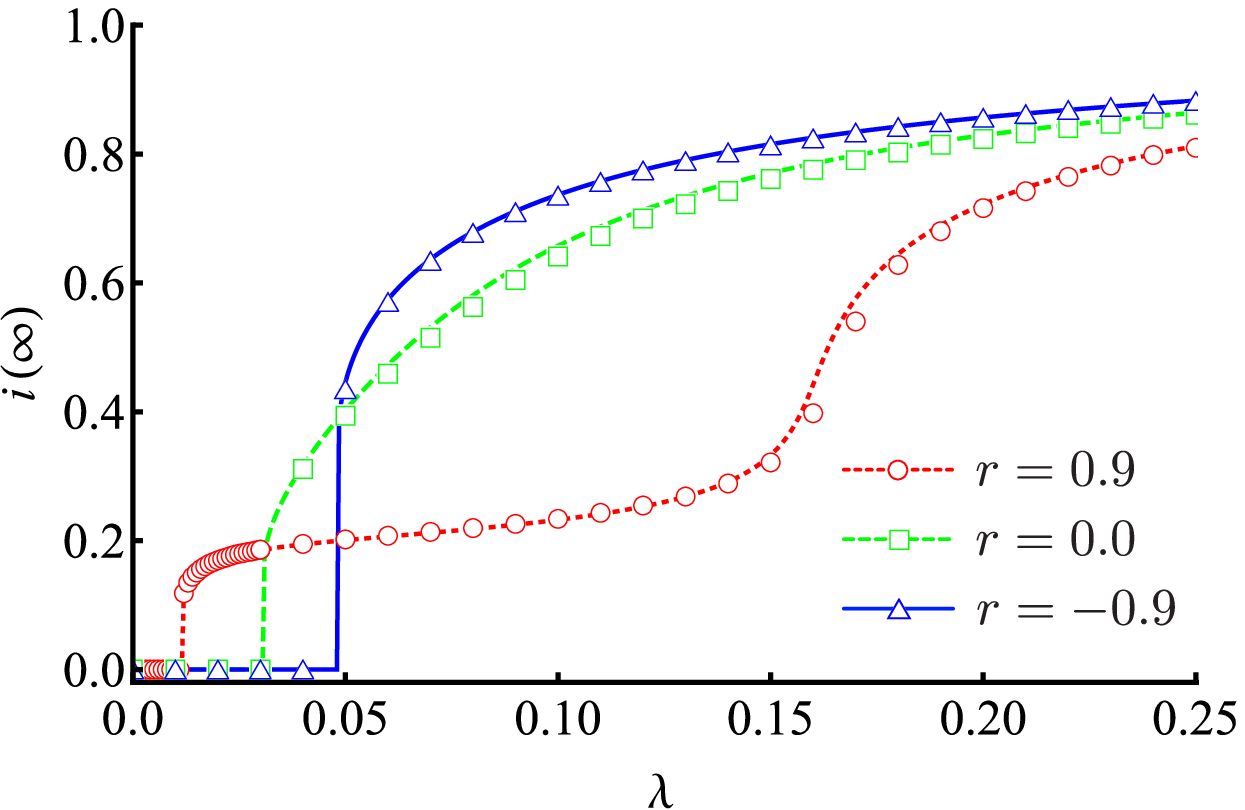}
(c)
\includegraphics[width=.30\textwidth]{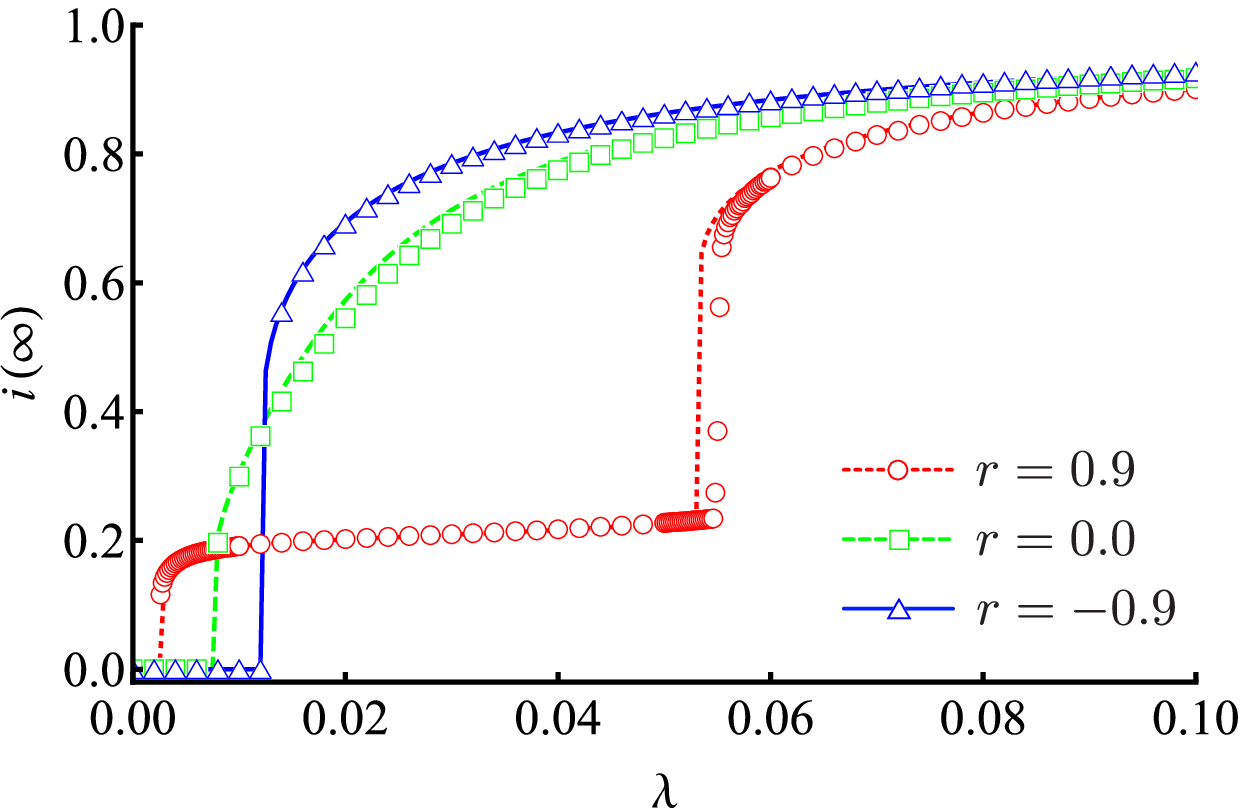}
\caption{
Comparisons of (quasi-)steady-state infected densities $i(\infty)$ determined in the AME calculations (lines) and Monte-Carlo simulations (symbols) in the synergistic SIS model on correlated bimodal networks with (a) $\alpha=0$, (b) $\alpha=1$, and (c) $\alpha=5$. 
The red-dotted lines (red circles), green-dashed lines (green squares), and blue-solid lines (blue triangles) represent the AME (Monte-Carlo) results of $r=0.9$, $r=0$, and $r=-0.9$, respectively.
The initial condition is set to $\rho=1$. 
}
\label{fig:infectedDensity-AME}
\end{center}
\end{figure}

\begin{figure}[t]
\begin{center}
(a)
\includegraphics[width=.30\textwidth]{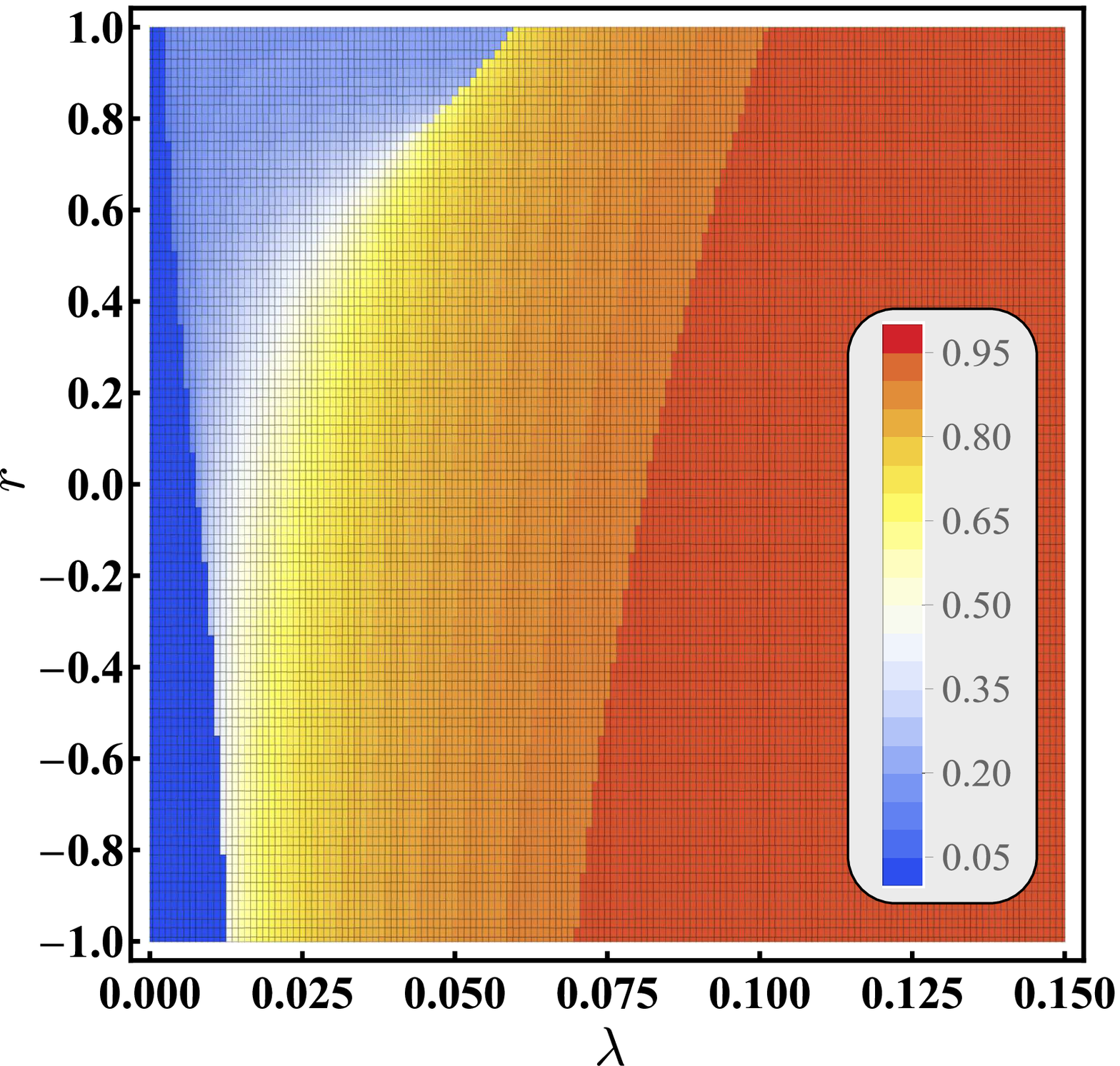}
(b)
\includegraphics[width=.30\textwidth]{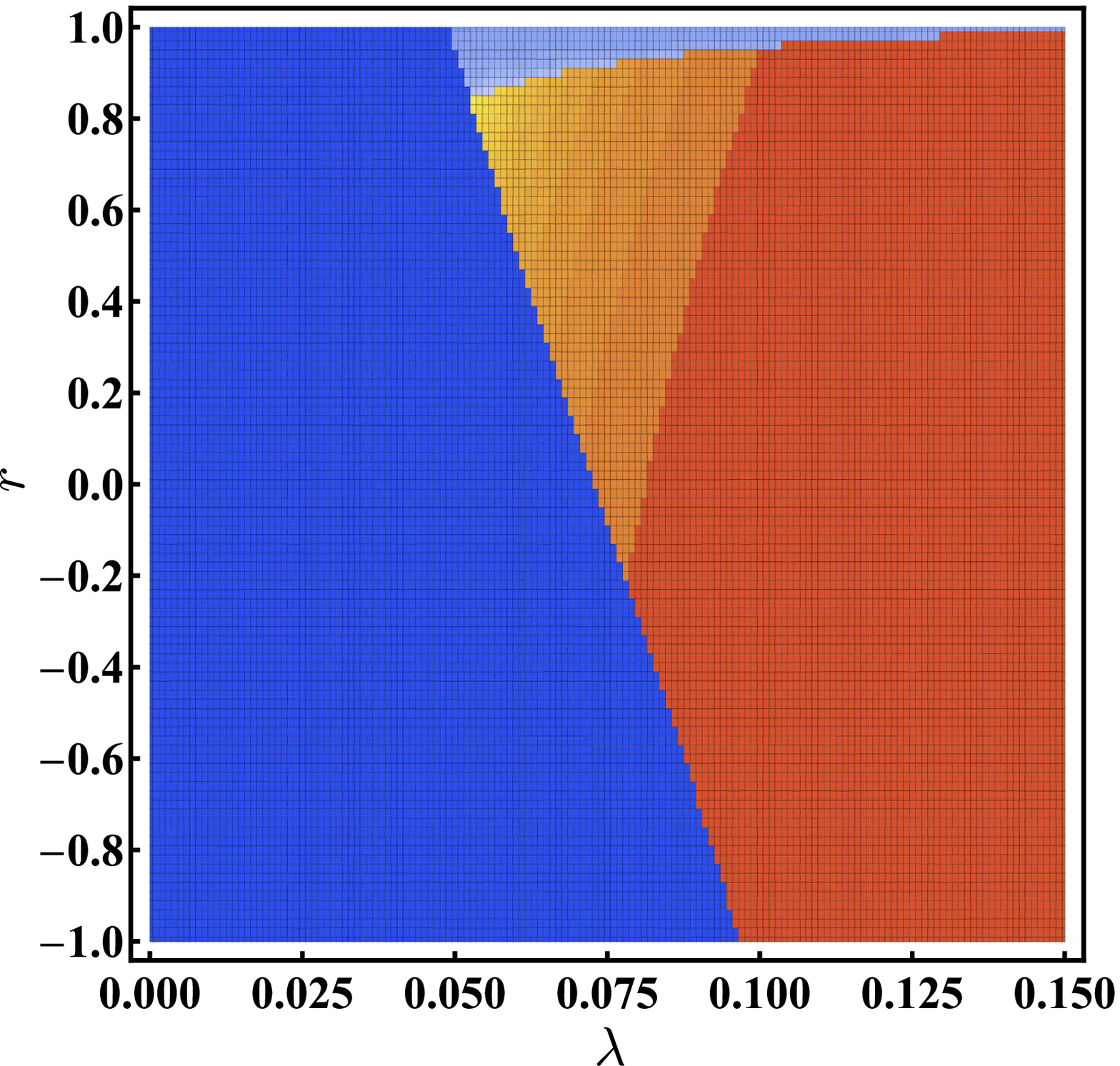}
(c)
\includegraphics[width=.30\textwidth]{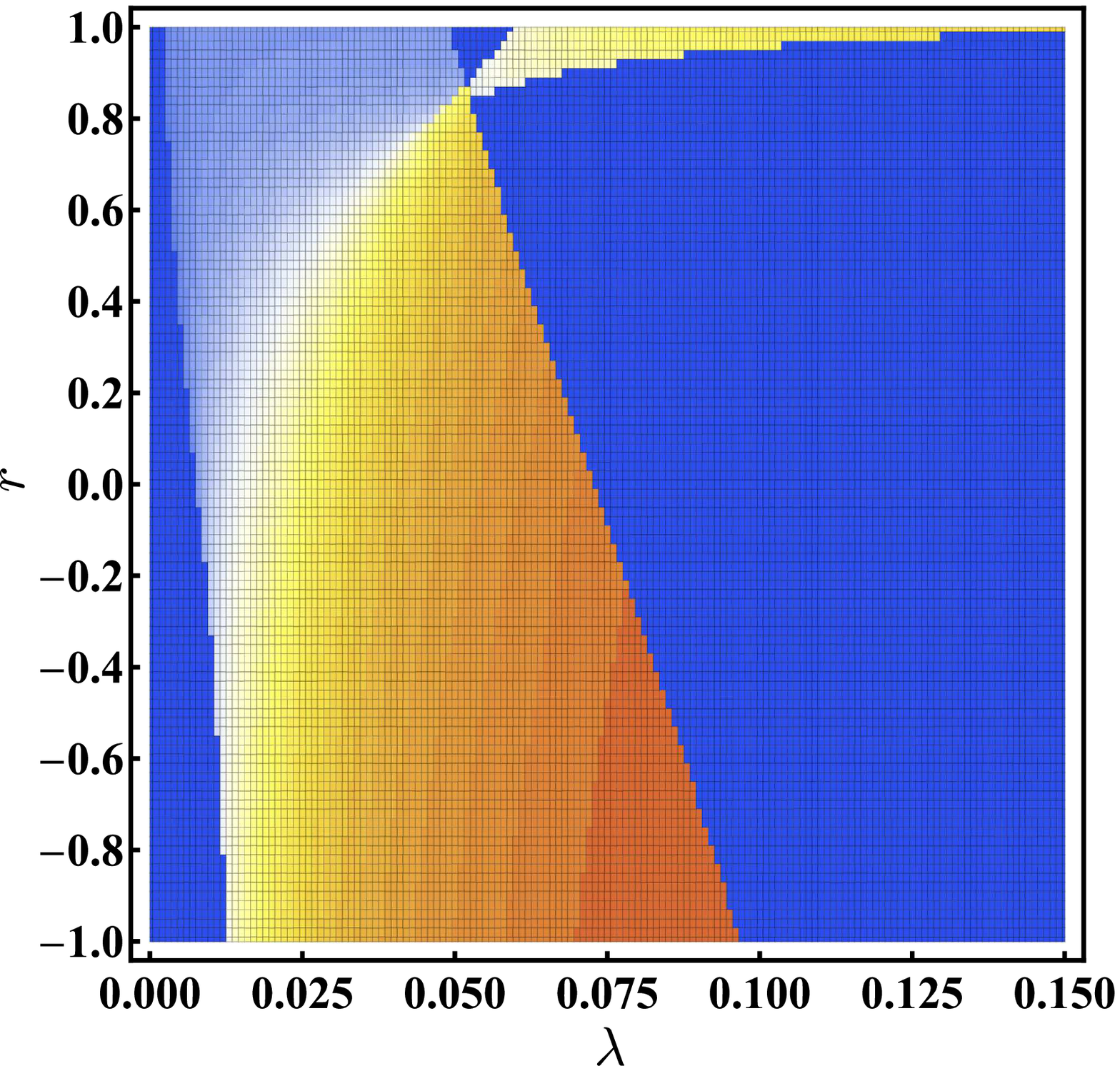}
\caption{
Color-maps of $i(\infty)$ in the ($\lambda,r$) plane for $\alpha=5$. 
The color-coded values represent (a) $i^{\rm (a)}$, the infected density $i(\infty)$ when $\rho=1$, (b) $i^{\rm (b)}$, the infected density $i(\infty)$ when $\rho=10^{-5}$, and (c) $i^{\rm (a)}-i^{\rm (b)}$.
Here $i(\infty)$ is evaluated from the AMEs.
}
\label{fig:colorMap-AME}
\end{center}
\end{figure}

Figure~\ref{fig:infectedDensity-AME} compares the $i(\infty)$'s obtained by the AME and Monte-Carlo simulations of the synergistic SIS model on the correlated bimodal networks with (a) $\alpha=0$,  (b) $\alpha=1$, and (c) $\alpha=5$.
In each panel, we employ correlated bimodal networks with $r=0.9$ (red-dotted line), $r=0$ (green-dashed line), and $r=-0.9$ (blue-solid line).
All simulations are performed on correlated bimodal networks with $N=10^4$.
The Monte-Carlo simulations begin with all nodes being infected ($\rho=1$ at time $t=0$) and continued until no infected nodes remains or until $t$ reaches $t_{\rm max}=10^2$.
For each parameter setting, we obtain $i(t_{\rm max})$ averaged over 100 runs.
We treat the simulation results of $i(t_{\rm max})$ (symbols) as the (quasi-)steady state infected density $i(\infty)$.
As shown in panels (a)--(c) of Fig.~\ref{fig:infectedDensity-AME}, the results of Monte-Carlo simulations well agree with those of the AME approach, confirming that the AMEs quantitatively describe the synergistic SIS model on the correlated bimodal networks.
Comparing Figs.~\ref{fig:infectedDensity:neutralBimodal}, \ref{fig:infectedDensity:assortativeBimodal}, and \ref{fig:infectedDensity-AME}, we also confirm that the mean-field prediction reproduces the continuity/discontinuity and other qualitative characters of the synergistic epidemics, although the results quantitatively deviates from the Monte-Carlo and AME results.
Both the Monte-Carlo simulations and AME calculations predict two discontinuous jumps when a strong synergistic effect works on a strongly assortative network (red-solid line in Fig.~\ref{fig:infectedDensity-AME} (c)), as earlier predicted by the mean-field analysis.
The AMEs also predict complex hysteresis behavior in correlated bimodal networks (Fig.~\ref{fig:colorMap-AME}).
Panels (a) and (b) of Fig.~\ref{fig:colorMap-AME} show the steady-state infected density $i(\infty)$ in the $(\lambda,r)$ plane when the synergy strength is $\alpha=5$ and the initial conditions are $\rho=1$ and $\rho=10^{-5}$, respectively, and Fig.~\ref{fig:colorMap-AME} (c) shows their difference.
The non-blue areas in Fig.~\ref{fig:colorMap-AME} (c) signify that $i(\infty)$ depends on $\rho$, confirming the emergence of hysteresis.

\section{Summary}

We have investigated the effect of degree correlation on spreading behaviors of synergistic epidemics through bimodal networks consisting of two node types with different degrees.
We have developed a mean-field treatment for the synergistic SIS model on such bimodal networks with a degree correlation.
Regardless of the synergy effect (absent or present), a positive and negative degree correlation in the SIS model diminishes and enlarges the epidemic threshold, respectively, affirming that infectious diseases invade and persist more easily on assortative networks than disassortative ones.
The synergistic SIS model undergoes a continuous transition when the synergy effect is absent or extremely weak and a discontinuous transition otherwise. 
On strongly assortative bimodal networks, and when the synergy effect is sufficiently strong, the SIS model exhibits two discontinuous jumps (explosive spreading events).
Nontrivial behaviors predicted by the mean-field treatment have been confirmed through extensive simulations and AMEs.

Two discontinuous jumps would be observed even though a network is degree uncorrelated.
For example, suppose that the type-1 and type-2 nodes in a bimodal network have the same degree, $k_1=k_2$, with keeping node type of each node assigned.
As the assortativity coefficient $r$ for node types of two ends of edges increases, the network is divided into two communities with the same degree. 
In a random initial configuration with $i_1(0)=i_2(0)=\rho$, at most, one explosive spreading can occur for any synergy strength and assortativity.
In contrast, when the initial state is seeded within only one community (e.g., $i_1(0)=\rho/p_1$ ($0<\rho \le p_1$) and $i_2(0)=0$), the strong synergistic effect and distinct community structure induce two explosive spreadings.
This is because the infection rate required for the persistence of epidemics is smaller than the one required for the invasion in the present model.

\begin{figure}[t]
\begin{center}
(a)
\includegraphics[width=.45\textwidth]{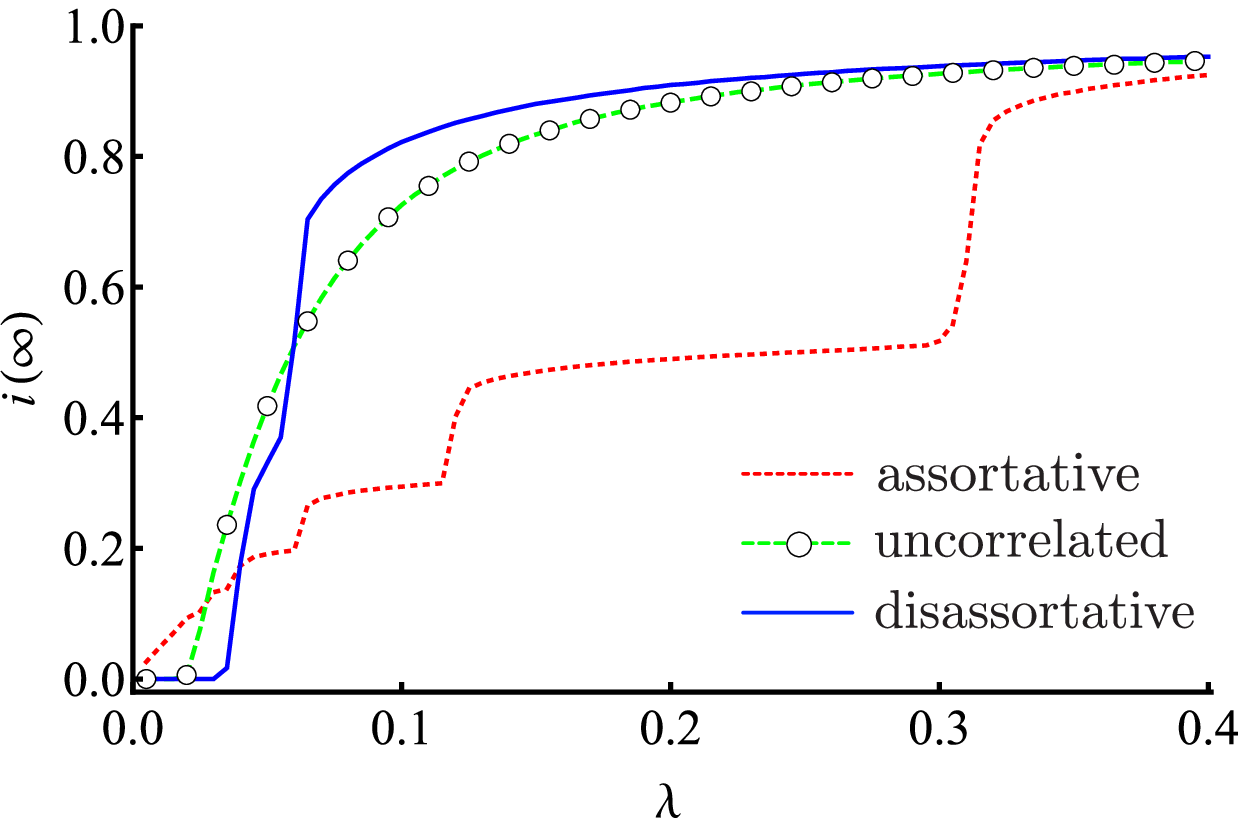}
(b)
\includegraphics[width=.45\textwidth]{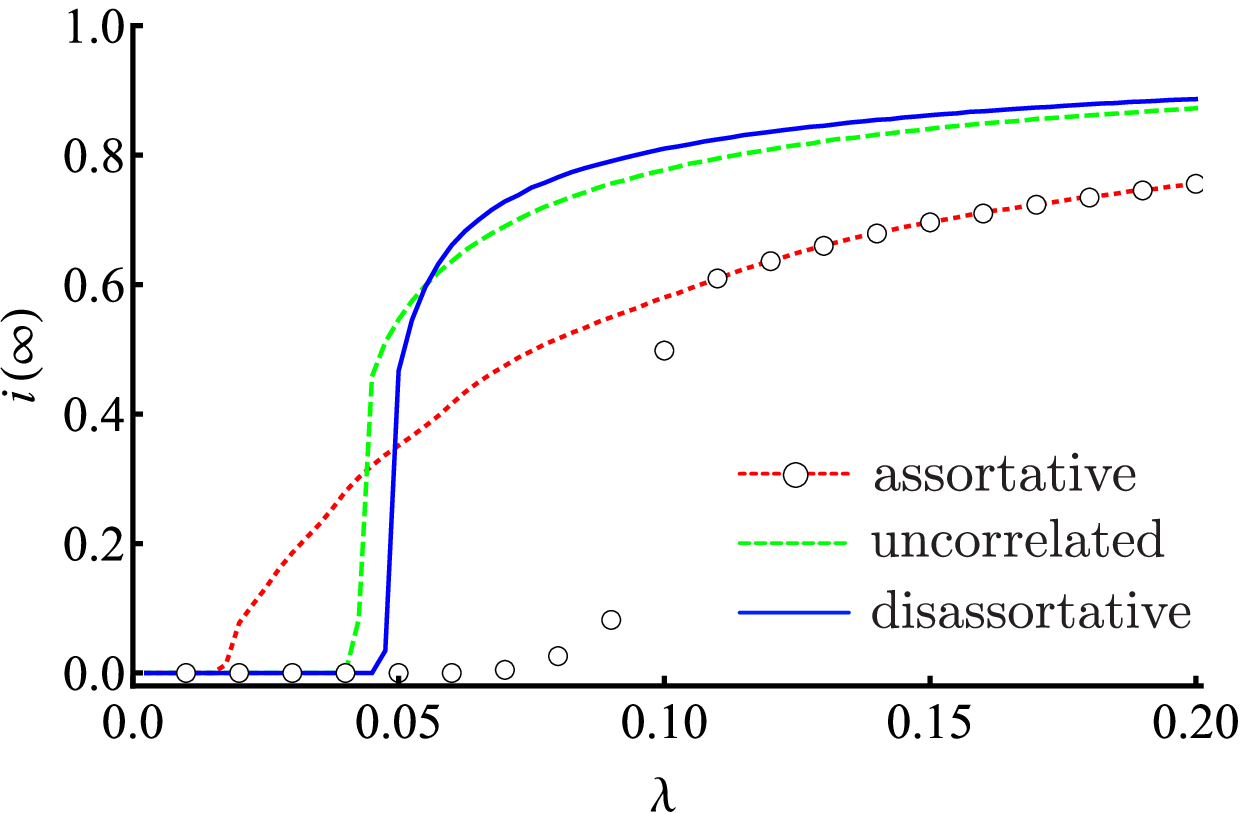}
\caption{
Monte-Carlo results for the synergistic SIS model with synergy strength $\alpha=5$ on (a) scale-free networks and (b) Erd\H{o}s-R\'{e}nyi networks with/without degree correlation. 
Scale-free networks for panel (a) have the degree distribution $P(k)\sim k^{-\gamma}$ with the degree exponent $\gamma=3$, the minimum degree $k_{\rm min}=3$, the maximum degree $k_{\rm max}=\sqrt{N}$, and the number of nodes $N=10^4$; Erd\H{o}s-R\'{e}nyi networks for panel (b) have the Poisson degree distribution with $\langle k \rangle = 5$ and $N = 10^4$ nodes.
To realize a correlated network, we first prepare an uncorrelated network and rewire edges based on the local-optimal algorithm \cite{valdez2011effect}. Using the algorithm, we generate assortative scale-free networks with $r=0.9$ and disassortative ones rewired $10^3N$ times for panel (a), and assortative Erd\H{o}s-R\'{e}nyi networks with $r=0.9$ and disassortative ones with $r=-0.9$ for panel (b).
The red-dotted, green-dashed, and blue-solid lines represent the steady-state infected density $i(\infty)$ for the case of assortative, uncorrelated, and disassortative networks, respectively when $\rho=1$.
The open black circles in panel (a) represent the result for the case of uncorrelated scale-free networks with the random initial configuration of $\rho=10^{-1}$; those in panel (b) represent the result for the case of assortative Erd\H{o}s-R\'{e}nyi networks with the initial state where nodes with $k = 2, 3$, and $4$ are infected and other nodes are susceptible.}
\label{fig:synergy-sis-on-sf}
\end{center}
\end{figure}

In this work, we have concentrated on a correlated bimodal network to investigate the effect of degree-correlated structure analytically. 
Empirical networks, in reality, are more heterogeneous. 
We briefly discuss the impact of degree heterogeneity on synergistic epidemic spreadings.
Figure \ref{fig:synergy-sis-on-sf} (a) shows the Monte-Carlo results for the synergistic SIS model on scale-free networks with degree exponent $\gamma=3$. 
For uncorrelated scale-free networks, both discontinuous jump and hysteresis seem to be suppressed due to the degree heterogeneity (green-dashed line and open black circles).
The data also show that the steady-state infected density $i(\infty)$ continuously increases from zero to a nonzero value, irrespective of degree correlation.
For strongly assortative networks (red-dotted line) and strongly disassortative networks (blue-solid line), however, the systems experience multiple discontinuous jumps of $i(\infty)$ at larger $\lambda$.
Strongly assortative networks consist of one comparatively-heterogeneous group (community) of large degree nodes and homogeneous groups of same degree nodes;
strongly disassortative networks consist of one heterogeneous group of large and small degree nodes and one comparatively-homogeneous group of similar degree nodes (whose degrees are neither too large nor too small) \cite{menche2010asymptotic}.
For both cases of assortative and disassortative networks, an epidemic is prevalent in the heterogeneous group containing large degree nodes by small infectivity, while the infectivity is too small for the infection to persist in other groups not containing large degree nodes.
It is natural to consider that multiple jumps of $i(\infty)$ at larger infection rates reflect an explosive spreading occurring within each of homogeneous groups.
Figure \ref{fig:synergy-sis-on-sf} (b) shows the Monte-Carlo results for the synergistic SIS model on the Erd\H{o}s-R\'{e}nyi networks with/without degree correlation. 
A discontinuous jump is observed for uncorrelated networks (green-dashed line) and disassortative networks (blue-solid line) in that degree variability is small. 
The data for assortative Erd\H{o}s-R\'{e}nyi networks of $\rho=1$ seem not to show a discontinuous jump of $i(\infty)$ (red-dotted line), whereas we confirm that a hysteresis behavior emerges when the initial state is adjusted (open black circles). 
A further study on synergistic epidemics on correlated networks having degree inhomogeneity should be needed in that  spreading behaviors depend crucially and sensitively on the initial state of the system.

The epidemic model in the present study is as simple as the classical SIS model and does not capture factual situations like the superspreading events of COVID-19. A recent paper proposed a model of superspreading in which an infectiousness heterogeneity is incorporated and clarified that the reduction in contact numbers is more important than that in contact time to mitigate epidemics with superspreaders \cite{nielsen2021covid}.
Future works are left to how superspreaders act in synergistic epidemics and how the number of contacts affects synergistic spreadings in degree-correlated networks.

\section*{Acknowledgment}
T.H.\ would like to thank Kanako Suzuki for their helpful comments.
S.M.\ and T.H.\ acknowledge the financial support from JSPS (Japan) KAKENHI Grant Number JP18KT0059.
T.H.\ acknowledges the financial support from JSPS (Japan) KAKENHI Grant Number JP19K03648.

\bibliography{ref-synergy}

\end{document}